\documentclass[fleqn,usenatbib]{mnras}
\usepackage{newtxtext,newtxmath}
\usepackage[T1]{fontenc}
\usepackage{amsmath}
\usepackage{epsfig}
\usepackage{algpseudocode}
\usepackage{algorithm}
\usepackage{xcolor}

\title[Treecode2]{Treecode2: The Power of Pluralism. I. Static Tests}

\author[J. E. Barnes]{Joshua E. Barnes\\
  Institute for Astronomy, University of Hawaii at Manoa\\
  2680 Woodlawn Drive, Honolulu, HI 96822, USA\\
  \textsf{barnes@hawaii.edu}}

\date{\small \today}

\pubyear{\the\year{}}

\begin{document}
\label{firstpage}
\pagerange{\pageref{firstpage}--\pageref{lastpage}}
\maketitle

\begin{abstract} \noindent
I describe an `oct-tree' $N$-body code which randomly shifts,
reorients, and resizes the root cell at each time step.  Averaging
over a plurality of root cell positions and orientations statistically
restores translational and rotational invariance.  The potentials and
forces which result can be much more accurate than those obtained from
a single force calculation.  In this paper, the principle of averaging
is tested on static configurations.  The next paper will show how this
technique can substantially improve global energy, momentum, and
angular momentum conservation at a negligible computational cost.
\end{abstract}


% \clerpage \newpage

\section{INTRODUCTION}
\label{sec:introduction}

An $N$-body code designed to simulate the evolution of a collisionless
self-gravitating system has two key components.  One evaluates the
mutual gravitational forces between $N$ particles; the other uses
these forces to step the velocities and positions of these particles
forward in time.

Large values of $N$ are necessary to accurately represent the
distribution functions of collisionless systems.  This places a high
premium on the computational efficiency of the code.  Integrating
trajectories is relatively inexpensive; the computing time required
scales as $O(N)$, since each particle is handled independently.
However, the most accurate and general method of calculating
gravitational forces involves evaluating the mutual interactions
between all \textit{pairs} of particles.  The cost of this
\textit{direct-sum} method scales as $O(N^2)$, and for any reasonable
value of $N$ it is prohibitive unless special-purpose hardware is
available.

Over decades, much effort has been invested in finding faster ways to
calculate, if only approximately, the mutual gravitational forces
between $N$ particles.  Two major types of algorithms have resulted
\citep[Ch.~2.9]{BT2008}.  \textit{Expansion} methods represent the
gravitational potential using a finite number of basis functions; for
example, Cartesian particle-mesh codes \citep[e.g.][]{HE1988} use
trigonometric functions, while spherical-harmonic codes
\citep[e.g.][]{M1984, HO1992} employ multipole expansions.  These
algorithms treat each particle independently, so their computing costs
scale as $O(N)$.  But expansion methods have significant constraints;
mesh codes have limited resolution, while
spherical-harmonic codes can only handle specific problem geometries.
Moreover, these schemes may violate galilean invariance; for example,
a spherical-harmonic code privileges the origin of the coordinate
system, and consequently does not conserve linear momentum.

\textit{Hierarchical} methods organize the particles into recursively
nested regions.  This organization is usually represented with a tree
structure \citep[][\S~2.3]{K1973}.  If the distance between two
regions is much greater than the diameter of either, the gravitational
field acting on one is insensitive to the details of the mass
distribution in the other.  This permits the forces between $N$
particles to be approximated at a cost scaling as $O(N \log N)$ or
even $O(N)$ \citep[e.g.][]{G1990}.  The accuracy of the resulting
forces can be adjusted by user-specified parameters, trading speed for
precision.  Hierarchical methods can achieve high spatial resolution
on arbitrary geometries, circumventing the restrictions of expansion
methods.

There are two ways to hierarchically organize the particles.
\textit{Lagrangian} methods \citep{A1981, A1985, J1985, P1985, P1986,
BBCP1990} typically adopt binary tree structures, in which each
interior node has exactly two descendents.  For example, \cite{P1986}
start by finding a pair of mutually nearest neighbors and replacing
them by a pseudo-particle representing their centre-of-mass; this
process is repeated until only a single pseudo-particle, identified as
the `root' of the tree, is left.

In contrast, \textit{Eulerian} methods (\citealt{BH1986}, hereafter
BH86; \citealt{GR1987}; \citealt{D2000}; \citealt{S2002}) begin by
enclosing the particles in a rectangular cell.  This root cell is then
recursively subdivided, e.g.~until each particle is isolated in a
separate sub-cell. The most widely used structure, known as an
`oct-tree', adopts cubical cells with up to eight descendents (BH86;
\citealt{GR1987}), each describing a separate octant.  Oct-tree-codes
are relatively easy to implement, lend themselves to at least some
level of analysis, and run fairly well on various kinds of hardware.

However, by defining a root cell with a specific position and
orientation, Eulerian tree-codes violate translational and rotational
invariance.  In concrete terms, the outcome of a treecode simulation
depends on the position and orientation of the system being simulated;
more abstractly, treecode force calculation does not commute with
translation or rotation operators.  These violations are `weak' in the
sense that they can be controlled by adjusting the accuracy
parameters, but this always entails a computational cost, and even
weak violations of galilean invariance may have consequences.

One such consequence is very easily demonstrated. Consider an isolated
system of particles; let $m_{p}$ and $\tilde{\mathbf{a}}_{p}$ be the
mass and tree-code acceleration of particle $p$, and define
\begin{equation}
  \tilde{\mathbf{F}}_\mathrm{bulk} =
    \sum_{p} m_{p} \tilde{\mathbf{a}}_{p} \, ,
\end{equation}
where the sum runs over all particles, to be the net or bulk force on
the system.  \cite{H1987} showed that tree-codes conserve linear
momentum imperfectly; in other words,
$\tilde{\mathbf{F}}_\mathrm{bulk}$ \textit{does not vanish}.  This is
a result of violating Newton's third law: the force a single particle
exerts on a distant cell is not \textit{exactly} equal and opposite to
the force the cell exerts on the particle.  Bulk forces \textit{per
  se} are not a result of violating translational invariance (see
\S~\ref{sec:discussion}).  But for an Eulerian code,
$\tilde{\mathbf{F}}_\mathrm{bulk}$ depends on the \textit{position} of
the system within the root cell; force calculation and translation
don't commute.

\begin{figure}
  % \centering
  \includegraphics[width=\linewidth]{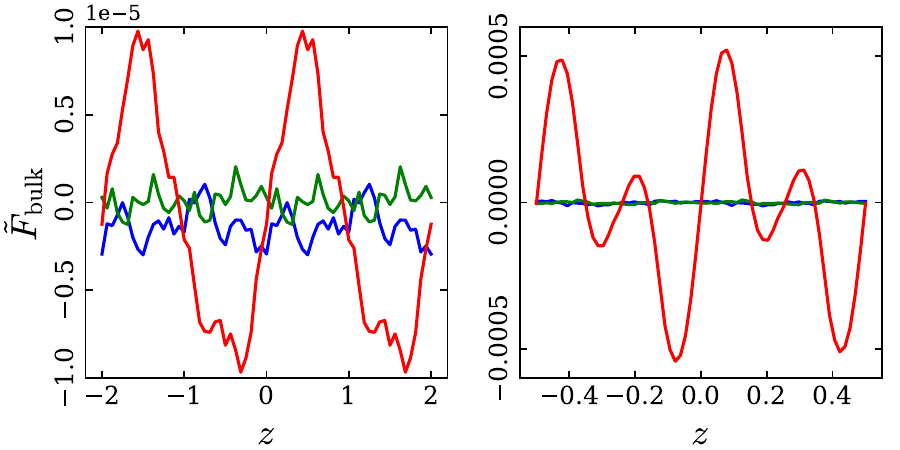}
  \caption{Bulk forces on spherical (left) and disc (right)
    configurations as functions of position. The sphere is a
    $3$-dimensional Gaussian with unit dispersion, while the disc is
    generated from the sphere by scaling $z$ coordinates by $0.1$;
    both are realized with $N=2^{18}$ particles.  Tree-code forces are
    computed with accuracy parameter $\theta=1$ as each configuration
    is stepped along the $z$-axis.  Blue, green, and red curves show
    the bulk forces in the $x$, $y$, and $z$ directions, respectively.
    Note that the vertical range of left-hand plot is $\pm 1.0 \times
    10^{-5}$, as indicated by the scaling factor at top, while the
    range of the right-hand plot is $60$ times greater.}
  \label{fig:plot_bulkforce_zcoord}
\end{figure}

Fig.~\ref{fig:plot_bulkforce_zcoord} illustrates this point.  The two
examples shown here use a root cell which is centred on the origin and
spans from $-8$ to $+8$ length units along the $x$, $y$, and $z$
directions.  The left-hand plot shows forces on a unit Gaussian sphere
as its centre is shifted from $z = -2$ to $z = 2$; $x$, $y$, and $z$
components are shown in blue, green, and red, respectively.  All three
components vary cyclically; the $z$ component has a much larger
amplitude than the other two.  Note that the origin an unstable point;
positive $z$ displacements yield positive $z$ forces, and vice versa.
The plot on the right shows bulk forces on a Gaussian disc with
flattening of $0.1$ along the $z$-axis.  As the disc translates along
its short axis, the maximum force in the $z$ direction is $\sim 50$
times larger than in the spherical case!  Indeed, pure discs simulated
with a tree-code sometimes `lift off', perpendicular to their spin
plane, unless force calculation is tuned to prioritize accuracy over
computing time.

The periodic nature of these forces reflects the structure the
oct-tree imposes on the representation of the mass distribution.  For
example, the trees which result before and after translating the
sphere in the left-hand panel of Fig.~\ref{fig:plot_bulkforce_zcoord}
by $2$ length units along the $z$ axis are almost identical;
consequently, the force errors which result are very similar.

The issue of momentum conservation has previously been addressed by
\cite{D2000}, who introduced an oct-tree algorithm which explicitly
symmetrizes gravitational sources and sinks.  This strictly enforces
Newton's third law; a single particle and a distant cell interacting
with each other will have equal and opposite forces.  Moreover,
Dehnen's code achieves $O(N)$ performance by using multipole
expansions, instead of lists of interactions, to represent the
gravitational field.

The code described here takes a different approach, with the goal of
mitigating violations of translational and rotational invariance by
averaging over the coordinate systems used to define the tree.  The
origin, size, and orientation of the root cell are randomly chosen at
each successive time-step; each resulting tree provides a different
representation of the mass distribution.  Individual force
calculations still violate invariance, but over time, translational
and rotational invariance are statistically recovered.  This
substantially improves the code's conservation of energy, linear
momentum, and angular momentum.  Moreover, force errors largely
decorrelate from one timestep to the next, delivering averaged forces
which are more accurate than the results of individual force
calculations.

A related method was described by \cite{W+2020}, who used cosmological
zoom-in simulations with periodic boundary conditions \citep{S2005} to
study the formation of extremely low-mass dark haloes.  To prevent
residual force-calculation errors from corrupting halo formation, they
randomly translated the particle configuration within the periodic
simulation volume at each timestep \citep{SPZR2021}.  For periodic
boundary conditions, this is equivalent to randomly translating the
origin of the oct-tree.  Notably, this technique was adopted
\textit{in addition} to Dehnen's algorithm; averaging over random
translations offers advantages beyond momentum conservation.

This paper presents static tests of the new force calculation
algorithm, deferring a suite of dynamical tests to a sequel (see
\S~\ref{sec:prelim_dynam_tests} for a preview).
\S~\ref{sec:methods} describes the new code, starting with the
rather simple changes needed to implement randomized tree coordinate
systems.  I also discuss some older modifications which improve
the code's accuracy, efficiency, and robustness.
\S~\ref{sec:static_tests} presents force-calculation tests for a
variety of mass distributions, characterizing errors in potential and
acceleration, and \S~\ref{sec:tree_averaging} shows how averaging can
reduce these errors.  Discussion and conclusions appear in
\S~\ref{sec:discussion} and \S~\ref{sec:conclusions}, respectively.  A
`modified' version \citep{B1990} of the code, designed to take
advantage of special-purpose hardware, is described in an appendix.

% \clerpage \newpage

\section{METHODS}
\label{sec:methods}

In outline, the new force-calculation algorithm is very similar to the
one described by BH86.  Given a set of $N$ particles with masses
$m_{p}$ and position vectors $\mathbf{r}_{p}$, where $p$ references
particles $1, \dots, N$, the code first fits a cubical cell around the
system.  This cell is then sub-divided into eight smaller cells;
subdivision continues recursively until each particle is isolated
within its own sub-cell.  The resulting hierarchy can be represented
by a directed graph of `nodes' (cells and particles) in the form of a
tree.  Each cell directly references (up to) eight nodes, known
collectively as the cell's `descendents'; on account of this eight-way
branching, this structure is called an `oct-tree'.

A typical tree has $O(\log N)$ levels\footnote{Pathological particle
distributions can generate much deeper trees, but in practice these do
not arise as Monte-Carlo representations of smooth distribution
functions.}, so the computing time required for tree construction is
$O(N \log N)$.  Each cell's total mass, centre-of-mass position, and
quadrapole moment are computed as part of the construction process.

The gravitational field at any point $\mathbf{r}$ can then be
evaluated by a partial recursive scan of the tree, starting from the
root cell.  Each cell encountered in this process is subject to a
simple test (\S~\ref{sec:opening_criteria}).  If it is sufficiently
distant from $\mathbf{r}$, then the field the cell generates is
approximated from its mass, position, and quadrapole moment;
conversely, if the cell is too close, it is `opened', meaning that
each of its immediate descendents is examined instead.  Particles
which are encountered during the scan are handled as regular two-body
interactions.  In effect, this procedure approximates the field at
$\mathbf{r}$ using a representation of the mass distribution which
captures nearby details but becomes increasingly coarse-grained at
larger distances.  Each level of the tree contributes a bounded number
of cells and particles to this representation, so the computing time
required to evaluate the field at $\mathbf{r}$ is $O(\log N)$.  If
forces are required on all $N$ particles, the computing time is $O(N
\log N)$.  Force calculation is significantly more expensive than tree
construction, but both obey the same scaling with $N$.

\subsection{Averaging Algorithm}
\label{sec:averaging_algorithm}

In the new code, the orientation, scale, and centre of the root cell
are chosen randomly.  To be specific, let $\mathsf{R}_{k}$ be a
rotation matrix chosen randomly at the $k^\mathrm{th}$
timestep\footnote{Strictly speaking, the static tests presented here
do not employ timesteps; $k$ is simply an integer. The timestep
terminology is used to simplify the description.}.  Likewise, let
$\mathit{S}_{k}$ be a random scale factor, drawn from a uniform
distribution in $\log\mathit{S}_{k}$ with $|\log\mathit{S}_{k}| \le
\log\mathit{S}_\mathrm{max}$.  Finally, let $\mathbf{T}_{k}$ be a
translation vector drawn from a uniform distribution within some
maximum radius $T_\mathrm{max}$.  Together, $\mathsf{R}_{k}$,
$\mathit{S}_{k}$, and $\mathbf{T}_{k}$ define an affine
transformation\footnote{Eq.~(\ref{eq:tree_coords}) preserves angles,
ensuring that cubes are cubes.  Non-cubical cells tend to have large
high-order moments, which may require octapole or higher corrections.}
from the coordinate system (CS) used by the simulation to the one used
in building the tree: if $\mathbf{r}_{p}$ is the simulation position
of particle ${p}$, its tree coordinate is
\begin{equation}
  \mathbf{r}^\prime_{p} =
    \mathsf{R}_{k} \mathit{S}_{k} (\mathbf{r}_{p} - \mathbf{T}_{k}) \, .
  \label{eq:tree_coords}
\end{equation}

Tree construction uses the transformed positions
$\mathbf{r}^\prime_{p}$ instead of $\mathbf{r}_{p}$.  The root cell is
centred on the origin of the tree CS; its size is the smallest power
of $2$ sufficient to enclose all particles.  This guarantees that the
coordinates of cell centres and boundaries can be accurately
represented as binary floating-point numbers; it also implies that
setting $S_\mathrm{max} = \sqrt{2}$ covers all functionally different
possibilities.  The resulting tree is very similar to a traditional
oct-tree, but organizes particles into a set of cells which are
randomly oriented, sized, and positioned with respect to the
simulation coordinates.  Calculation of cell properties -- masses,
centre-of-mass positions, and quadrapole moments -- can be carried out
in simulation coordinates, and force calculation follows the usual
hierarchical algorithm.

This technique depends critically on randomizing $\mathsf{R}_{k}$,
$\mathit{S}_{k}$, and $\mathbf{T}_{k}$, with no correlation between
timesteps.  Holding these quantities fixed provides no benefit; it
simply replaces one preferred CS with another.  But when a
\textit{different} tree CS is chosen at each timestep, force errors at
timestep $k$ will partly cancel at timestep $k+1$. Over many
timesteps, no CS is preferred, and the algorithm should gradually
approach translational and rotational invariance.

\subsection{Quadrapole Moments}
\label{sec:quadrapole_moments}

Following \citet{H1987}, the code uses the quadrapole moments of cells
to improve the accuracy of gravity calculations.  Consider a cell $c$
containing particles $p \in c$ with masses $m_{p}$ and positions
$\mathbf{r}_{p}$; the cell has total mass $m_{c}$ and centre of mass
position $\mathbf{r}_{c}$.  Let $\mathbf{s}_{p} = \mathbf{r}_{p} -
\mathbf{r}_{c}$ be the position of particle $p$ relative to the cell's
centre of mass; then the cell's traceless quadrapole moment is
\begin{equation}
  \mathsf{Q}_{c} =
    \sum_{p \in c} m_{p} (3 \, \mathbf{s}_{p} \!\otimes\! \mathbf{s}_{p}
                       - s_{p}^2 \mathsf{I}) \, ,
  \label{eq:quadrapole_moment}
\end{equation}
where $\mathsf{I}$ is the identity matrix.  The cell's gravitational
potential at position $\mathbf{r} = \mathbf{s} + \mathbf{r}_{c}$ can
then be expanded to quadrapole order:
\begin{equation}
  \phi_{c}(\mathbf{s}) \simeq
    - \frac{G \, m_{c}}{s}
    - \frac{G \, \mathbf{s} \!\cdot\! \mathsf{Q}_{c} \!\cdot\! \mathbf{s}}%
           {2 s^5} \, .
  \label{eq:quadrapole_potential_nosoft}
\end{equation}

This expression should be modified to include Plummer softening
\citep{A1963}.  In the limiting case where $c$ contains a single
particle, so that all higher-order moments vanish, the softened
potential is
\begin{equation}
  \phi_{c}(\mathbf{s}) =
    - \frac{G \, m_{c}}{(s^2 + \epsilon^2)^{1/2}} \, ,
\end{equation}
where $\epsilon$ is the softening length.  By analogy, \citet{H1987}
implemented softening by replacing $s$ with $(s^2 + \epsilon^2)^{1/2}$
in the denominators of \textit{both} terms in
eq.~(\ref{eq:quadrapole_potential_nosoft}).  This gives acceptable
results in the small-$\epsilon$ limit, but is somewhat less accurate
in situations where large amounts of mass are concentrated on scales
of order $\epsilon$ \citep{B2012}.

A better approximation to the softened quadrapole potential was
suggested by Keigo Nitadori (personal communication, 2015):
\begin{equation}
  \phi_{c}(\mathbf{s}) \simeq
    - \frac{G \, m_{c}}{(s^2 + \epsilon^2)^{1/2}}
    - \frac{G \, (\mathbf{s}\!\cdot\!\mathsf{Q}_{c}\!\cdot\!\mathbf{s}
    - \epsilon^2 \tilde{Q}_{c})}{2 (s^2 + \epsilon^2)^{5/2}} \, ,
  \label{eq:quadrapole_potential}
\end{equation}
where
\begin{equation}
  \tilde{Q}_{c} = \sum_{p \in c} m_p \, s_p^2 \, .
\end{equation}
This approximation substantially improves the accuracy of softened
body-cell interactions, while imposing only a modest computational
cost.  It is implemented as a compile-time option. 

The acceleration due to the cell is $\mathbf{a}_{c} = - \partial
\phi_{c} / \partial \mathbf{r}$.

\subsection{Opening Criteria}
\label{sec:opening_criteria}

To decide if a cell $c$ is so close to a particle $p$ that it
\textit{must} be opened, BH86 used a simple geometrical test.  Let $s
= |\mathbf{r}_\mathrm{p} - \mathbf{r}_\mathrm{c}|$ be the distance
between $p$ and $c$, and let $\ell_c$ be the edge-length of $c$.  Then
$c$ must be opened if
\begin{equation}
  s < \ell_\mathrm{c} / \theta \, ,
  \label{eq:bh86_criterion}
\end{equation}
where $\theta$ is a parameter, of order unity, which controls the
accuracy of the force calculation.  In principle, this geometrical
criterion insures that the maximum relative error from any individual
cell is bounded \citep[e.g.,][]{BH1989}.  This test can be implemented
very efficiently; the value of $(\ell_\mathrm{c} / \theta)^2$ is
stored in each cell during tree construction, and can be quickly
compared to $s^2$ for every cell encountered during a recursive tree
scan.

In practice, this simple criterion can fail if the centre-of-mass (CM)
of a cell is far from its geometric centre.  For example, if the CM
lies in one corner of $c$ and $p$ lies at the opposite corner, then
eq.~(\ref{eq:bh86_criterion}) will fail unless the parameter $\theta <
1 / \sqrt{3} \simeq 0.577$.  Noticing this, \cite{H1987} considered an
additional test to make sure that $c$ does not contain $p$, although
experiments with a limited set of particle distributions suggested
this was unnecessary provided that $\theta \lesssim 1.2$.
Subsequently, however, \cite{SW1994} showed that
eq.~(\ref{eq:bh86_criterion}) could fail catastrophically for $\theta
\sim 1$ if $p$ is replaced with a compact, gravitationally bound
system of particles; in this situation, interactions within the bound
system are not evaluated, resulting in the `detonating satellite'
phenomenon.

To cure this problem, I take the offset $d_c$ between $c$'s geometric
centre and its CM into account \citep{B1994}.
Eq.~\ref{eq:bh86_criterion} is replaced with
\begin{equation}
  s < \ell_\mathrm{c} / \theta + d_c \, .
  \label{eq:offset_criterion}
\end{equation}
This criterion positively guarantees that any cell which contains $p$
will be opened provided that $\theta < 2 / \sqrt{3} \simeq 1.16$,
which is larger than the $\theta$ values typically employed.
Heuristically, eq.~(\ref{eq:offset_criterion}) gives `extra attention'
to cells which have large offsets $d_c$ relative to their sizes; this
seems reasonable since such cells are likely to have significant
high-order moments.  The new code uses
eq.~(\ref{eq:offset_criterion}), although the original criterion is
available as a run-time option.

Any tree scan which correctly opens every cell containing particle $p$
will \textit{necessarily} visit $p$ itself.  This provides a simple
check which is easily implemented in the force-calculation tree scan.
By default, a tree scan which fails this test triggers a fatal error.

\subsection{Threading the Tree}
\label{sec:threading_the_tree}

Since force calculation entails performing many tree scans, it is
important to make tree traversal as efficient as possible.  During
construction, each cell of the tree stores the addresses of its
immediate descendents in an ordered array.  The resulting
data-structure can only be traversed by a recursive algorithm, or
equivalently, by an iterative algorithm which maintains an explicit
stack of tree nodes (particles and cells) to process.  This imposes a
significant overhead on each tree scan. But as \cite{M1990a} noted, a
straightforward transformation permits the tree to be scanned with a
simple iterative algorithm.

The basic idea of Makino's transformation is familiar to
anyone who has perused a hierarchically-structured document.  Each
page of information offers two hyperlinks: one goes to a
page which discusses the \textit{next} topic, while the other (which
is optional) goes to a page or pages which discuss the
present topic in \textit{more} detail.

Algorithm~\ref{code:iterative_tree_scan} presents the iterative tree
scan.  Here $n$ is either a cell or a particle, and $s$ is a flag
which indicates if the scan has correctly visited $p$.  Each cell $c$
stores two pointers: $\mathsf{Next}(c)$ is the node to be visited if
$c$ does not need to be opened, while $\mathsf{More}(c)$ is the first
descendent of $c$.  Each particle $p$ just stores one pointer,
$\mathsf{Next}(p)$.  The constant $\mathsf{ROOT}$ is the root cell of
the tree, and $\mathsf{NULL}$ is a sentinel value marking the end of
the tree.  The function $\mathsf{IsCell}(n)$ is true if $n$ is a cell,
and $\mathsf{MustOpen}(n,p)$ implements the test in
eq.~(\ref{eq:offset_criterion}).

\begin{algorithm}[t!]
  \caption{Force calculation for particle $p$ by iterative tree scan.}
  \begin{algorithmic}
    \State {$n \gets \mathsf{ROOT}$}
    \State {$s \gets \mathsf{FALSE}$}
    \While {$n \ne \mathsf{NULL}$}
      \If {$\mathsf{IsCell}(n)$}
        \If {$\mathsf{MustOpen}(n,p)$}
          \State {$n \gets \mathsf{More}(n)$}
        \Else
          \State {process interaction of $p$ with cell $n$}
          \State {$n \gets \mathsf{Next}(n)$}
        \EndIf
      \Else
        \If {$n \ne p$}  
          \State {process interaction of $p$ with particle $n$}
        \Else
          \State {$s \gets \mathsf{TRUE}$}
        \EndIf
        \State {$n \gets \mathsf{Next}(n)$}
      \EndIf
    \EndWhile
    \If {$s \ne \mathsf{TRUE}$}
      \State {error: scan did not visit $p$}
    \EndIf
  \end{algorithmic}
  \label{code:iterative_tree_scan}
\end{algorithm}

The tree is threaded with $\mathsf{Next}$ and $\mathsf{More}$ pointers
following the initial phase of tree construction.  Once the tree is
threaded, the eight descendent pointers of a cell are no longer
needed, and the memory they occupy can be reused, in particular, for
quadrapole moments.

In passing, it's worth mentioning that the $\mathsf{Next}$ pointer is
also useful during the initial phase of tree construction.  On
occasion, two particles may have \textit{exactly} the same positions,
at least to $32$-bit precision\footnote{This typically arises when a
system with a high central density, such as a \cite{J1983} model, is
displaced a few scale lengths from the origin.}.  This situation is
easily detected as particles are inserted into the tree structure; if
it occurs, the second particle is linked after the first using the
$\mathsf{Next}$ pointer.  This linkage is preserved when the tree is
threaded.

% \clerpage \newpage

\section{STATIC TESTS}
\label{sec:static_tests}

While the value of tree averaging should ultimately be demonstrated by
fully dynamical simulations (Paper~II), the results of such
simulations are hard to interpret without a good understanding of the
errors arising at each time step.  Therefore, this paper presents
tests using static configurations.  The test procedure is described in
\S~\ref{sec:test_procedure}.  I then examine errors in force
calculations for individual particles(\S~\ref{sec:single_particles}),
error distributions for all particles
(\S~\ref{sec:error_distributions}), trends with opening angle
(\S~\ref{sec:errors_vs_opening_angle}), the origin of bulk forces and
torques (\S~\ref{sec:bulk_forces} and \ref{sec:bulk_torques}), and
computing times (\S~\ref{sec:computing_time}).  In these tests, I
employ a single tree CS, replacing eq.~(\ref{eq:tree_coords}) with
$\mathbf{r}'_{p} = \mathbf{r}_{p}$.  The goal here is to establish a
baseline with which to evaluate the effects of tree averaging.

An emerging theme of this study is that tree-code force errors are
\textit{not statistically independent}.  This is not a new observation
\citep[e.g.,][]{BH1989}, but its implications have not always been
recognized.  An understanding of error correlations is valuable in
interpreting tree-code results.

\subsection{Test Procedure}
\label{sec:test_procedure}

The particle configurations used to test the algorithm are naturally
organized into a $5 \times 5 \times 5$ array.  The first dimension of
this array selects one of five different mass models.  The second
dimension fixes the number of particles used to realize each model,
ranging in multiplicative steps of $4$ from from $N = 2^{14} = 16384$
to $N =2^{22} = 4194304$ (using powers of two for $N$ helps ensure
that particle masses and their sums are accurately represented as
floating point numbers).  The third dimension specifies a unique
random number seed used when constructing each specific
realization.  The five realizations constructed for each choice of
mass model and $N$ are statistically equivalent to each other; by
comparing results across an ensemble of realizations, one can gain
some idea of statistical uncertainties.

The five mass models include three spherical systems with different
density profiles, a flattened disc, and a small group of spherical
systems, as follows:
\begin{enumerate}

  \item[(i)] A \cite{H1990} model, with mass $M=1$ and scale radius
    $a=1$.  The density profile is smoothly tapered at radius $b=100$
    \citep{B2012}. The model is offset from the origin by a random
    vector $\mathbf{r}_\mathrm{off}$ drawn from a uniform
    distribution, with magnitude $|\mathbf{r}_\mathrm{off}| \le 4$
    length units.

  \item[(ii)] A \cite{J1983} model, with mass $M=1$ and scale radius
    $a=1$.  The density profile is smoothly tapered at radius
    $b=100$. The model is randomly offset.

  \item[(iii)] An \cite{E1965} model, with mass $M=1$, half-mass
    radius $r_\mathrm{h}=1$ and index $n=3.5$.  The model is randomly
    offset.

  \item[(iv)] A flattened disc, with mass $M=1$, generated by drawing
    coordinates $(x_{p},y_{p},z_{p})$ from a unit Gaussian, and then
    rescaling $z_{p} \to 0.1 z_{p}$.  The disc is randomly offset
    \textit{and} rotated.

  \item[(v)] A group of $4$ \cite{E1965} models, each generated with
    $N/4$ particles, and offset to positions chosen at random within a
    sphere of radius $4$.

\end{enumerate}
Random offsets and/or rotations are used to examine how the tree-code
behaves when the distribution of particles is not aligned with the
coordinate system.  Again, a different seed was used each time these
were chosen.

A mass model with density $\rho(\mathbf{r})$ is realized by choosing
$N$ positions $\mathbf{r}_{p}$, where $p$ labels particles $1, \dots,
N$.  The chance of choosing position $\mathbf{r}$ is directly
proportional to $\rho(\mathbf{r})$; consistent with this strategy,
each particle has equal mass $m_{p} = M / N$.  Positions are drawn
independently, so the number of particles in any given volume obeys
Poisson statistics.  This gives rise to `microscopic' differences
between the configurations in a given ensemble; in contrast, the
random offsets and/or rotations introduce `macroscopic' differences.

For each realization, I first computed reference forces and potentials
via a simple direct-sum algorithm, setting $G = 1$ and using Plummer
softening with $\epsilon = 0.01$.  Direct-sum forces take an
excessively long time to evaluate for $N > 2^{20}$, so for
realizations with $N = 2^{22}$, I computed forces on a subset of
$2^{18}$ particles which uniformly sample the full system.  Let
$\phi_{p}$ and $\mathbf{a}_{p}$ be the resulting potential and
acceleration for particle $p$.  These quantities were computed in
double precision to avoid roundoff issues with sums of $O(N)$ terms.

I then used the tree-code to compute approximate potentials
$\tilde{\phi}_{p}$ and accelerations $\tilde{\mathbf{a}}_{p}$ for each
particle $p$ in all $125$ realizations.  By default, all test
calculations included quadrapole moments (results obtained without
them are briefly discussed in \S~\ref{sec:errors_vs_opening_angle}).
Forces were calculated for $7$ opening-angle values equally spaced
from $\theta = 0.4$ to~$1.0$, using the same set of realizations for
each.  This strategy introduces a subtle complication: a realization
which happens to be an outlier for one $\theta$ value is often an
outlier for other values as well.  It would have been cleaner to use a
unique set of realizations for each $\theta$ value, but the direct-sum
forces were too expensive.  Single precision was adequate for this
range of $\theta$ values, since each sum contains only $O(\log N)$
terms.

The basic quantities of interest are the net or absolute potential and
acceleration errors for particle $p$:
\begin{equation}
  \Delta \tilde{\phi}_{p} \equiv
    \tilde{\phi}_{p} - \phi_{p} \, ,
  \qquad
  \Delta \tilde{\mathbf{a}}_{p} \equiv
    \tilde{\mathbf{a}}_{p} - \mathbf{a}_{p} \, .
  \label{eq:def_abs_error}
\end{equation}

\subsection{Single Particles}
\label{sec:single_particles}

A tree-code approximates the potential and acceleration of particle $p$
by summing its interactions with a set $\mathsf{P}_{p}$ of particles
and another set $\mathsf{C}_{p}$ of cells
(eq.~\ref{eq:quadrapole_potential}).  The former are, by definition,
accurate, but each of the latter is an approximation which introduces
some error.  Let $\tilde{\phi}_{p,c}$ and $\tilde{\mathbf{a}}_{p,c}$
be tree-code approximations for the potential and acceleration of cell
$c$, while $\phi_{p,c}$ and $\mathbf{a}_{p,c}$ are computed by
direct-sum over all particles in $c$.  Then the errors for cell $c$
are $\Delta \tilde{\phi}_{p,c} \equiv \tilde{\phi}_{p,c} - \phi_{p,c}$
and $\Delta \tilde{\mathbf{a}}_{p,c} \equiv \tilde{\mathbf{a}}_{p,c} -
\mathbf{a}_{p,c}$.  These cell-by-cell errors can be summed to obtain
the absolute errors:
\begin{equation}
  \Delta \tilde{\phi}_{p} =
    \sum_{c \in \mathsf{C}_{p}} \Delta \tilde{\phi}_{p,c} \, ,
  \qquad
  \Delta \tilde{\mathbf{a}}_{p} =
     \sum_{c \in \mathsf{C}_{p}} \Delta \tilde{\mathbf{a}}_{p,c} \, .
  \label{eq:sum_grav_error}
\end{equation}

\begin{figure}
  \centering
  \includegraphics[width=\linewidth]{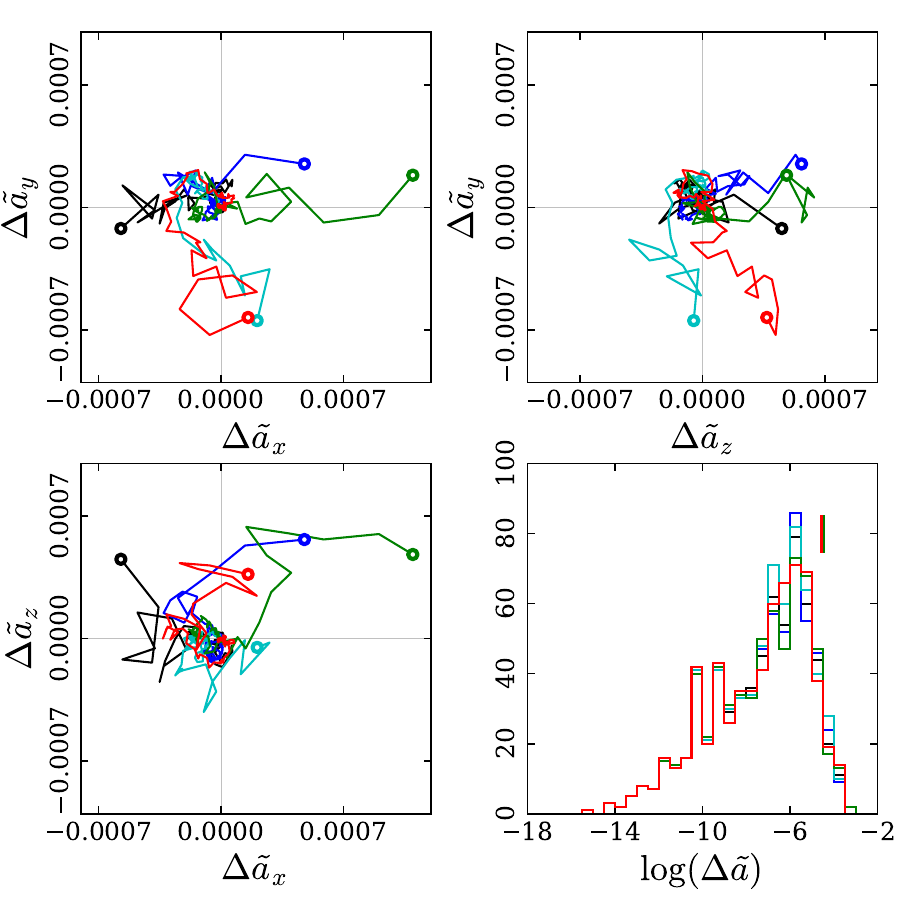}
  \caption{Acceleration errors for single particles, plotted as
    quasi-random walks.  Five particles were sampled from radius $r
    \simeq 1$ in a Hernquist model; each is plotted in a different
    color.  The model has $N = 2^{18}$ particles, and forces were
    computed with $\theta = 1.0$.  Top left, top right, and bottom
    left panels show projections of acceleration errors on the
    $(x,y)$, $(z,y)$, and $(x,z)$ planes, respectively.  Steps are
    ordered from smallest to largest in length; an open circle marks
    the end of each walk.  Bottom right panel shows error
    distributions; vertical marks indicate RMS values of $\Delta
    a_{p,c}$.}
  \label{fig:plot_gravwalk_3D_hernq}
\end{figure}

The sums in eq.~(\ref{eq:sum_grav_error}) can be visualized as
quasi-random walks.  Fig.~\ref{fig:plot_gravwalk_3D_hernq} presents
examples of $\Delta \tilde{\mathbf{a}}_{p}$ for a Hernquist model with
$N = 2^{18} = 262144$, computed using $\theta = 1.0$.  There's no
natural order to the set $\mathsf{C}_{p}$, so in these plots the steps
are ordered by increasing $|\Delta \tilde{\mathbf{a}}_{p,c}|$.

All of these random walks are dominated by their largest steps.  In
the three orthogonal views of the acceleration errors, a handful of
steps at the end account for most of the displacement from the origin.
The distributions of $\log \, |\Delta \tilde{\mathbf{a}}_{p,c}|$
plotted in the lower right panel support this; the vast majority of
cells each particle interacts with generate numerically small errors,
while a tiny minority yield large errors.  Although the cell-opening
criterion limits the \textit{relative} errors of cells in
eq.~(\ref{eq:sum_grav_error}), there is no corresponding bound on the
\textit{absolute} error.  It's not unusual for the total gravitational
force at a given point to be dominated by a relatively small number of
cells -- indeed, that's basically the point of a tree-code!  The
relative errors due to these dominant cells are typical, but their
absolute errors are still large.

Close inspection of Fig.~\ref{fig:plot_gravwalk_3D_hernq} hints that
acceleration errors due to different cells are \textit{correlated}
with each other.  None of these paths are linear, but most of them
seem to make fairly consistent progress away from the origin as they
reach their terminal points. However, the very uneven distribution of
step lengths makes it difficult to be certain.

\begin{figure}
  \centering
  \includegraphics[width=0.5\linewidth]{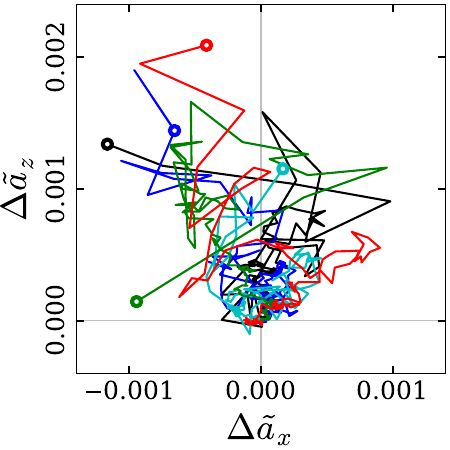}
  \caption{Acceleration errors for single particles from a disc model.
    Five particles were sampled from $\sqrt{x^2 + y^2} < 0.2$ and $z
    \simeq 0.15$.  Other details as for
    Fig.~\ref{fig:plot_gravwalk_3D_hernq}}
  \label{fig:plot_gravwalk_xz_disk}
\end{figure}

Correlations are obvious in Fig.~\ref{fig:plot_gravwalk_xz_disk},
which plots acceleration errors for the Gaussian disc used in
Fig.~\ref{fig:plot_bulkforce_zcoord}.  Here, the disc was positioned
at the origin, and forces for a sample of five particles drawn from
$\sqrt{x^2 + y^2} < 0.2$ and $z \simeq 0.15$ are plotted.  While
individual steps retain substantial random components, all five paths
are clearly biased in the positive $z$ direction, and this is
reflected in steps of all sizes.  Tree-codes systematically
underestimate the vertical acceleration fields of thin discs.  This
bias arises because highly stratified mass distributions generate
cells with significant octapole and higher (hereafter `octapole-plus')
moments, which are excluded when cell fields are truncated at
quadrapole order (\S~\ref{sec:quadrapole_moments}).

\subsection{Error Distributions}
\label{sec:error_distributions}

The absolute errors $\Delta \tilde{\phi}_{p}$ and $\Delta
\tilde{\mathrm{a}}_{p}$ for particle $p$ are fundamental quantities,
but in many cases it's useful to discuss the corresponding normalized
or \textit{relative} errors,
\begin{equation}
  \delta \tilde{\phi}_{p} \equiv
    \Delta \tilde{\phi}_{p} / |\phi_{p}| \, ,
  \qquad
  \delta \tilde{a}_{p} \equiv
    |\Delta \tilde{\mathbf{a}}_{p}| / |\mathbf{a}_{p}| \, .
  \label{eq:def_rel_error}
\end{equation}
Relative errors simplify comparison of results from different models,
and relative acceleration errors are more relevant in determining the
accuracy of particle trajectories.  Note, however, that particles at
the very centre of a model may yield alarmingly large values of
$\delta \tilde{a}_{p}$, since the normalizing acceleration
$|\mathbf{a}_{p}|$ can be very small.

\begin{figure*}
  \centering
  \includegraphics[width=\linewidth]{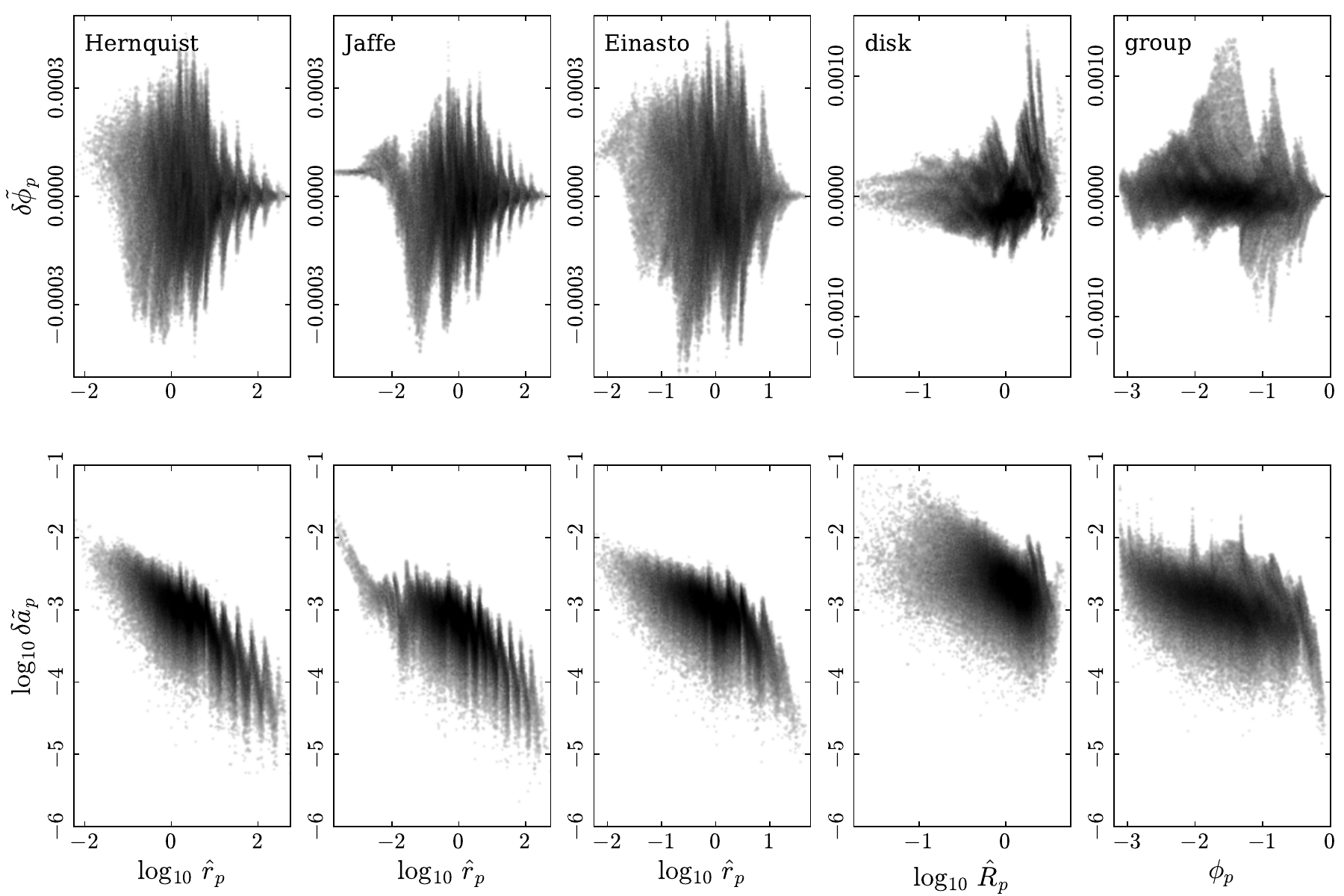}
  \caption{Particle-by-particle errors versus distance from centre or
    potential.  The top row plots relative potential error, while the
    lower one plots log relative acceleration error.  Forces were
    calculated using opening angle $\theta = 0.8$; quadrapole terms
    were corrected for softening, with $\epsilon = 0.01$.  Each
    realization has $N = 2^{18}$ bodies.}
  \label{fig:scat_err_rad}
\end{figure*}

Fig.~\ref{fig:scat_err_rad} presents scatter-plots of relative errors
in potential and acceleration.  For the three spherical models, the
horizontal axis shows the distance $\hat{r}_{p} = |\mathbf{r}_{p} -
\mathbf{r}_\mathrm{off}|$ from the system's centre, while for the disc
model, the cylindrical radius $\hat{R}_{p}$ is used instead.  The
group model has no unique centre, so the horizontal axis shows the
potential $\phi_{p}$.  Each realization of $N = 2^{18}$ particles is
represented by two panels: the upper one shows the relative potential
error $\delta \tilde{\phi}_{p}$, which must be plotted on a linear
scale, and the lower one shows the relative acceleration error $\delta
\tilde{a}_{p}$, plotted on a logarithmic scale.  The softening length
was $\epsilon = 0.01$.  Tree-code potentials and accelerations were
calculated using an opening angle $\theta = 0.8$; cell potentials
include softening-corrected quadrapole terms
(eq.~\ref{eq:quadrapole_potential}).  Unless otherwise stated, these
parameters are used in all subsequent force-calculation plots.

The errors for the three spherical models all display broadly similar
patterns.  In general, the structure of the errors reflects the
underlying oct-tree; the `barbs' in the $\delta \tilde{\phi}_{p}$
distributions are often spaced by factors of $2$ in $\hat{r}_{p}$.
This also appears in the `sawtooth' pattern of acceleration errors
$\log_{10} \, \delta \tilde{a}_{p}$, along with a general increase
toward the centre of each model.  At first glance, potential errors
may appear symmetric about $\delta \phi = 0$, but a second look
corrects this impression; the Jaffe model (top row, second panel)
provides a particularly clear example of asymmetry, with
$\tilde{\phi}_{p} > \phi_{p}$ everywhere within $\log_{10} \hat{r} <
-2$.

Errors for the disc and group models paint a different picture.  They
are significantly larger than those typical of spherical models.  The
fairly orderly structures of $\delta \tilde{\phi}_{p}$ versus
$\hat{r}_{p}$ seen in the spherical cases are replaced with much more
chaotic patterns.  Acceleration errors generally increase toward the
centre of the disc, or deeper within the potential well of the group,
but have wide distributions everywhere.  Potential errors are not
symmetrically distributed about $\delta \tilde{\phi} = 0$, with
$\tilde{\phi}_{p}$ overestimated in some places, and underestimated in
others.

\begin{figure}
  \centering
  \includegraphics[width=\linewidth]{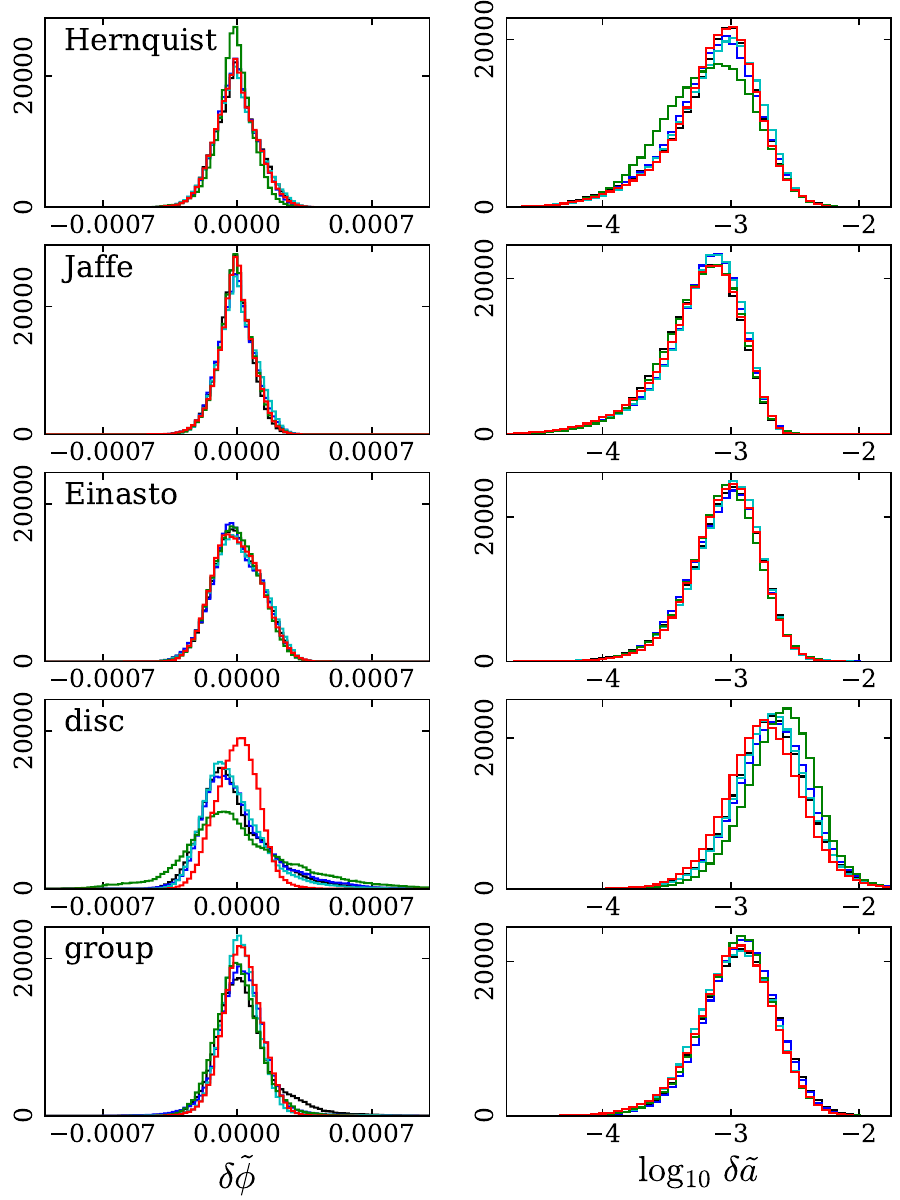}
  \caption{Distributions of relative potential errors (left) and log
    relative acceleration errors (right) for all five models.  Forces
    were calculated as in Fig.~\ref{fig:scat_err_rad}.  A different
    color is used for each realization.}
  \label{fig:hist_err}
\end{figure}

These scatter-plots display just one realization for each model.  Do
the other four ensemble members, each with its own offset, orientation
(for the disc model), layout (for the group model), and Monte-Carlo
sampling, yield similar results?  Fig.~\ref{fig:hist_err} plots
histograms of relative errors in potential $\delta \tilde{\phi}_{p}$
and acceleration $\log_{10} \, \delta \tilde{a}_{p}$ for all five mass
models.  Each realization is represented by a different color to help
show variations within each ensemble.

The Hernquist model (top row) yields distributions of $\delta
\tilde{\phi}$ which are nearly symmetric around zero, with only modest
variations between configurations.  Most of the variation is driven by
random offsets; Monte-Carlo sampling plays a much smaller role.  In
particular, the realization represented by the green histograms is
situated in a `sweet spot' with respect to the oct-tree.  As a result,
its $\delta \tilde{\phi}$ distribution (left panel) is narrower and
more peaked than the others; moreover, its $\log_{10} \delta
\tilde{a}$ distribution (right panel) is shifted toward smaller
values.

These sweet spots -- offsets which yield atypically accurate
potentials and forces -- are observed with all three spherical mass
models, although none of the other realizations shown in
Fig.~\ref{fig:hist_err} happens to fall near one.  One such spot
always lies at the origin (zero offset: $\mathbf{r}_\mathrm{off} = (0,
0, 0)$), but similar results arise with other offsets as well.  In
Fig.~\ref{fig:hist_err}, the realization giving rise to the green
histograms had an offset of $\mathbf{r}_\mathrm{off} = (-0.155,
-0.207, -3.935)$, adjacent to a sweet spot at $\mathbf{r}_\mathrm{off}
= (0, 0, 4)$.  In general, it appears that sweet spots result when the
bulk of the mass distribution is more or less evenly divided between
eight cells (which need not all be siblings).

In contrast, the disc models (fourth row) yield $\delta \tilde{\phi}$
distributions which vary widely and are manifestly \textit{asymmetric}
about $0$.  The potential and acceleration errors are substantially
larger than those found for spherical systems, consistent with
Fig.~\ref{fig:scat_err_rad}.  Again, most of the variation among the
ensemble members arises because of macroscopic differences in disc
orientation and position, as opposed to microscopic differences in
Monte-Carlo sampling.  The disc represented by the red histograms,
which happens to be tilted by only $\sim 5.3^\circ$ with respect to the
$x$--$z$ plane, yields significantly more accurate potentials and
accelerations compared to other ensemble members.  This is a further
consequence of singling out a specific coordinate system: discs which
are aligned with the coordinate planes of the tree are handled more
accurately.  Thus, in addition to sweet spots, flattened systems also
have `sweet orientations'.

The group realizations yield fairly narrow $\delta \tilde{\phi}$
distributions -- indeed, they somewhat narrower and more symmetric
than the distributions for the Einasto models making up these groups.
These distributions display noticeable variations arising from the
random construction of each realization.  However, these variations
are not reflected in the $\delta \tilde{a}$ histograms, which seem
remarkably consistent.  The acceleration errors are only slightly
larger than those obtained for individual Einasto models.

\subsection{Errors vs.~Opening Angle}
\label{sec:errors_vs_opening_angle}

\begin{figure}
  \centering
  \includegraphics[width=\linewidth]{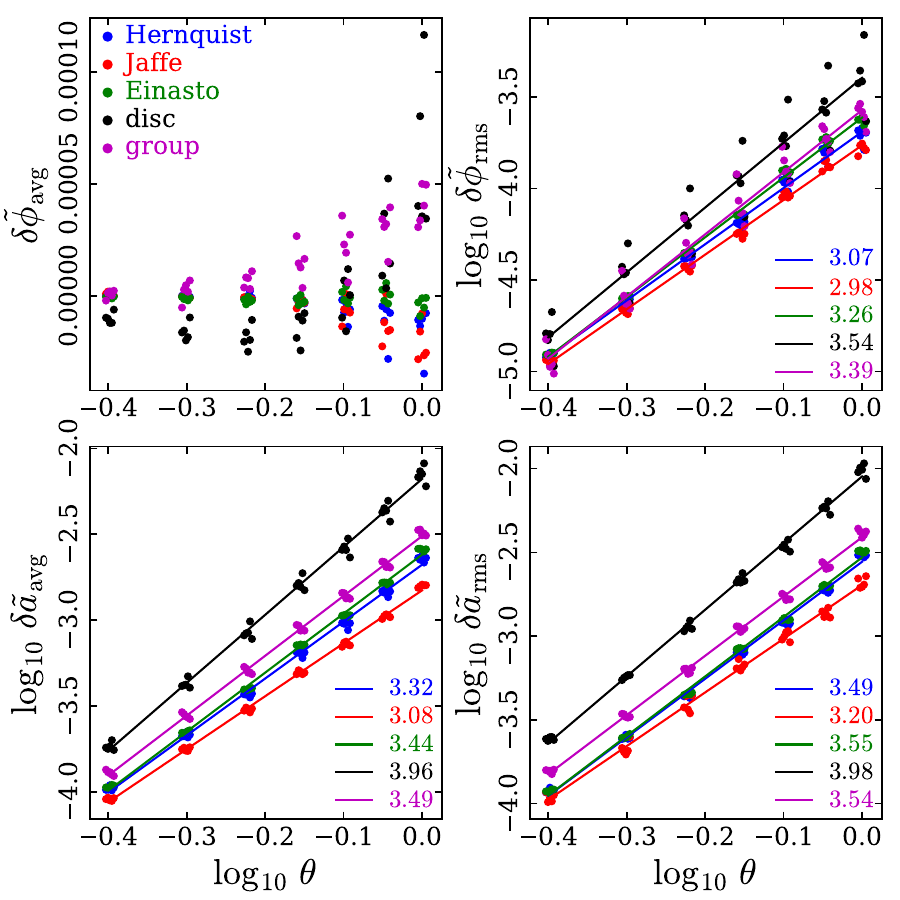}
  \caption{Mean and RMS errors versus log opening angle.  Top row
    shows potential errors, bottom shows acceleration; left column
    shows means, right shows RMS.  Forces were calculated as in
    Fig.~\ref{fig:scat_err_rad}.  Each model is represented by five
    realizations of $N = 2^{18}$ bodies; $\log_{10} \theta$ values are
    plotted with small offsets to reduce crowding.  The upper-left
    panel uses a linear $y$-axis since $\delta
    \tilde{\phi}_\mathrm{avg}$ may have either sign.  The other three
    panels employ log-log scaling; thin lines are least-squares fits,
    with the slopes listed in the lower right.}
  \label{fig:scat_err_theta}
\end{figure}

How do potential and acceleration errors depend on the opening angle?
To address this question, it helps to quantify the overall errors of a
force calculation with a few key numbers:
\begin{align}
  \delta \tilde{\phi}_\mathrm{avg} &\equiv
    \frac{1}{N} \sum_{p} \delta \tilde{\phi}_{p} \, , &
  \delta \tilde{\phi}_\mathrm{rms} &\equiv
    \sqrt{\frac{1}{N} \sum_{p} (\delta \tilde{\phi}_{p})^2} \, ,
  \nonumber \\
  \delta \tilde{a}_\mathrm{avg} &\equiv
    \frac{1}{N} \sum_{p} \delta \tilde{a}_{p} \, , &
  \delta \tilde{a}_\mathrm{rms} &\equiv
    \sqrt{\frac{1}{N} \sum_{p} (\delta \tilde{a}_{p})^2} \, ,
  \label{eq:def_agg_error}
\end{align}
Fig.~\ref{fig:scat_err_theta} shows how these errors depend on opening
angle $\theta$ for all five mass models.  The five realizations of
each model are plotted separately; each has $N = 2^{18}$ particles.

The upper left panel of this figure shows the relationship between
$\theta$ and the mean relative potential error $\delta
\tilde{\phi}_\mathrm{avg}$, which measures the net bias away from
$\delta \tilde{\phi} = 0$ in Fig.~\ref{fig:hist_err}.  Since this bias
may have either sign, the vertical axis uses a linear scale.  It's
evident that biases decrease as $\theta \to 0$, but the manner of
convergence depends on the mass model.  The three spherical models
seem to be slightly biased toward $\delta \tilde{\phi}_\mathrm{avg} <
0$ for $\theta \ge 0.7$ but converge rapidly as $\theta$ decreases.
On the other hand, the group models, each composed of four spherical
components with random positions, are biased toward $\delta
\tilde{\phi}_\mathrm{avg} > 0$ and converge more slowly.  Finally, the
disc models, with random positions \textit{and} orientations, exhibit
dramatic variations from one realization to the next; these models
appear to be biased toward $\delta \tilde{\phi}_\mathrm{avg} > 0$ for
$\theta > 0.8$, but the other way for $\theta < 0.8$.

The upper right panel of Fig.~\ref{fig:scat_err_theta} plots RMS
relative potential errors $\delta \tilde{\phi}_\mathrm{rms}$, which
effectively measure the spread around $\delta \tilde{\phi} = 0$.
Since the spread is positive-definite, logarithmic scaling can be used
on the vertical axis.  In general, the RMS errors are rather similar
for all five models, although the disc and group models yield larger
errors for a given $\theta$, and exhibit larger variations within
their respective ensembles.  This plot shows that $\delta
\tilde{\phi}_\mathrm{rms}$ follows rough power-law relationships with
$\theta$.  The color-coded lines give the slopes; all five models are
fairly well fit by $\delta \tilde{\phi}_\mathrm{rms} \propto
\theta^{(3.3 \pm 0.3)}$.

The two bottom panels plot average (left) and RMS (right) relative
acceleration errors.  These plots are quite similar, implying that the
error distributions are not dominated by outliers.  Consistent with
previous results, disc and group models produce the largest
acceleration errors, although variations among members of the disc and
group ensembles are smaller than those seen for potential
errors. Acceleration errors fall rapidly as $\theta$ is reduced.
Power-law fits yield $\delta \tilde{a}_\mathrm{rms} \propto
\theta^{3.4 \pm 0.2}$ for the three spherical models; the disc models
converge even faster, with $\delta \tilde{a}_\mathrm{rms} \propto
\theta^{4.0}$.

For completeness, Supplemental Fig.~1 shows how errors vary with
$\theta$ when forces are computed \textit{without} quadrapole terms.
In general, there's little reason not to use quadrapole terms
\citep{H1987}; the saving in computing time is relatively modest,
while the additional errors incurred are quite large.  The disc model
in particular is handled very poorly without quadrapole terms, because
a good part of the long-range force is due to cells which sample
substantial pieces of the disc; these cells will have large quadrapole
moments which are neglected in Supplemental Fig.~1.  For the Hernquist
and Einasto models, acceleration errors scale roughly as $\delta
\tilde{a}_\mathrm{rms} \propto \theta^{2.4 \pm 0.3}$; the Jaffe model
converges a bit more slowly.

By way of comparison, \cite{M1990b} derived large-$N$ scaling laws of
the form $\delta \tilde{a}_\mathrm{rms} \propto \theta^{7/2}$ (with
quadrapole moments) or $\delta \tilde{a}_\mathrm{rms} \propto
\theta^{5/2}$ (without).  This calculation assumed a homogeneous
particle distribution, implying that the quadrapole and octapole
moments of cells are due entirely to Poissonian fluctuations
\citep{BH1989}, and that separate cells are statistically independent.
The power laws reported above are broadly consistent with Makino's
predictions.  However, this agreement may be partly fortuitous; trees
constructed for inhomogeneous models contain cells which sample
gradients or large-scale features in the mass distribution, and these
cells have octapole-plus moments which are not necessarily dominated
by Poissonian noise.

Potential and acceleration errors gradually decrease as $N$ increases
\citep{H1987}.  Supplemental Fig.~2 illustrates the relationship
between $N$ and $\delta \tilde{a}_\mathrm{rms}$, computed using
opening angle $\theta = 0.8$.  For the values of $N$ used here,
acceleration errors follow rough power-laws, $\delta
\tilde{a}_\mathrm{rms} \propto N^\beta$, with indices in the range
$\beta \simeq -0.07$ to~$-0.15$ (summarizing for $\theta \in
[0.4,0.8]$).  As $N$ grows, the cells contributing to the field
of a typical particle are increasingly well-populated; this
should reduce the stochastic part of their octapole-plus moments
\citep{BH1989} and therefore drive down the overall error.  If
this explanation is correct, force errors should level off in
the $N \to \infty$ limit.

\subsubsection{Softening corrections}
\label{sec:softening_corrections}

\begin{figure}
  \centering
  \includegraphics[width=\linewidth]{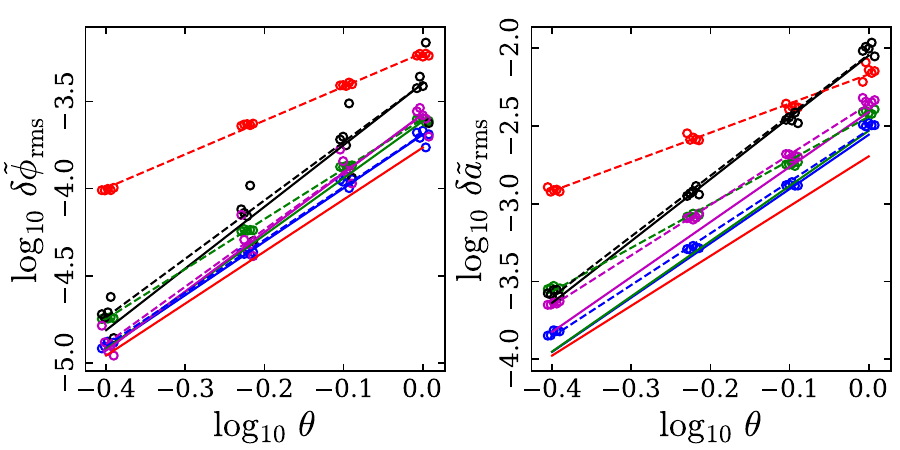}
  \caption{RMS potential (left) and acceleration (right) errors
    vs.~opening angle.  Open markers and dashed lines were computed
    \textit{without} the softening correction to the quadrapole term,
    while solid lines, copied from Fig.~\ref{fig:scat_err_theta},
    include this correction.}
  \label{fig:scat_err_theta_nc}
\end{figure}

Fig.~\ref{fig:scat_err_theta_nc} shows how the quadrapole softening
correction (eq.~\ref{eq:quadrapole_potential}) improves force
calculations.  For the softening length $\epsilon = 0.01$ adopted
here, the Jaffe model (red) is the only one which exhibits a large
effect.  This model contains a non-trivial amounts of mass ($\sim
1$~per cent) within a radius $r = \epsilon$ of its centre.  The
correction dramatically improves the accuracy of the calculations; for
$\theta = 1$, the corrected potential and acceleration are $\sim 3$
times more accurate, while for $\theta = 0.4$, they are a factor of
$\sim 10$ better.  Without the correction, the distributions of
$\delta \tilde{\phi}$ and $\delta \tilde{a}$ have tails \textit{much}
larger than typical errors.  These arise from particles within a
radius $\hat{r} \sim 10 \epsilon$ from the centre
\citep[Appendix~B]{B2012}, and account for the dramatic effects seen
here.

Results for the other mass models also improve, albeit some by only a
few per cent.  The correction yields essentially \textit{no} benefit
for Hernquist and disc models. In these models, cells within distances
of a few $\epsilon$ contribute relatively little to the gravitational
field; consequently, the net effect of the errors associated with
these cells is small.  Of course, this depends on the softening
length; a rough generalization is that softening corrections are
likely to be important if a few per cent of the total mass lies within
a sphere of radius $\epsilon$ -- as is the case for the Jaffe model.

\subsection{Bulk Forces}
\label{sec:bulk_forces}

Newton's third law implies that the total force on an isolated system
always vanishes: $\mathbf{F}_\mathrm{bulk} = \sum m_{p} \mathbf{a}_{p}
= 0$.  As Fig.~\ref{fig:plot_bulkforce_zcoord} demonstrates, forces
computed using a tree-code don't obey this law.  The bulk force
generated by a tree-code is the sum of the force errors for individual
particles:
\begin{equation}
  \tilde{\mathbf{F}}_\mathrm{bulk} \equiv
    \sum_{p} m_{p} \tilde{\mathbf{a}}_{p} =
      \sum_{p} m_{p} \Delta \tilde{\mathbf{a}}_{p} \, .
  \label{eq:bulk_force}
\end{equation}
If the errors $\Delta \tilde{\mathbf{a}}_{p}$ were statistically
independent, the second sum in eq.~\ref{eq:bulk_force} could be
treated as a random walk.  The magnitude of the bulk force could then
be estimated as $\tilde{F}_\mathrm{bulk} \simeq \Delta
\tilde{F}_\mathrm{rms} \sqrt{N}$, where $\Delta
\tilde{F}_\mathrm{rms}$ is the RMS force error.  However,
\S~\ref{sec:single_particles} has already presented a situation in
which the force errors on adjacent particles are clearly correlated.
As a result, bulk forces are generally much larger than a random-walk
model would predict.

\begin{figure}
  \centering
  \includegraphics[width=\linewidth]{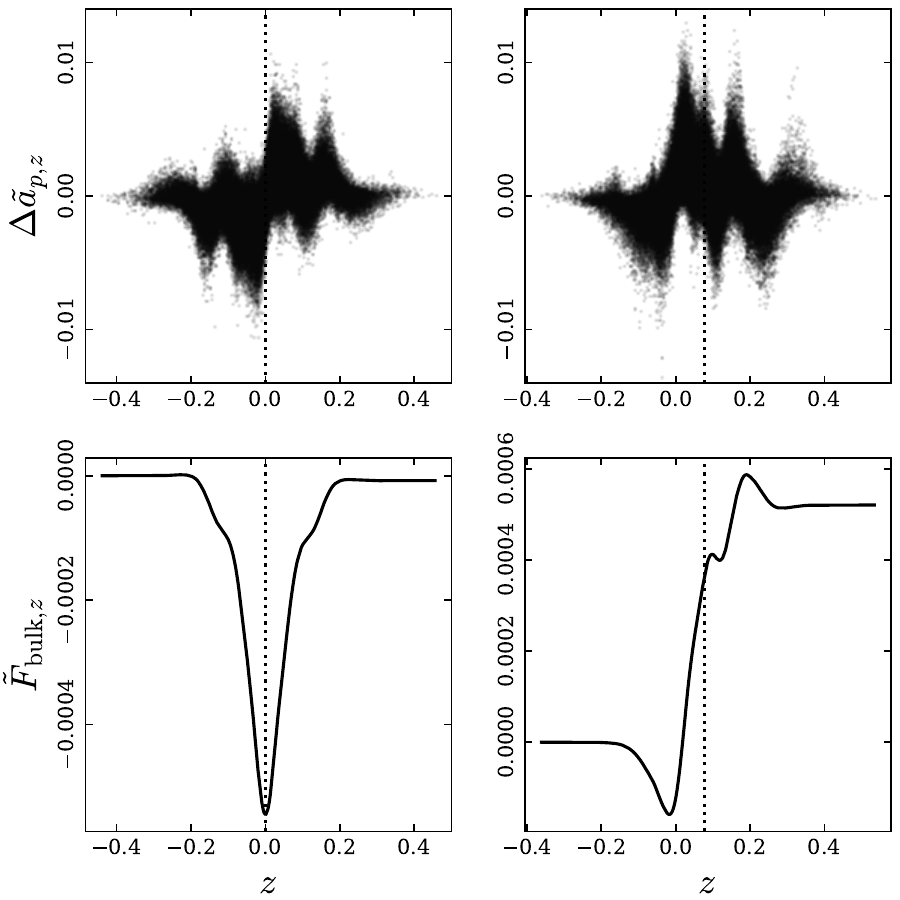}
  \caption{Origin of bulk forces on a disc system.  Top panels are
    scatter-plots of $z$-component acceleration errors vs.~$z$; bottom
    panels show $z$-component of bulk force for all particles with
    positions $<z$.  Dotted line shows the position of the disc's
    midplane.}
  \label{fig:plot_bulk_disk_zcoord}
\end{figure}

Fig.~\ref{fig:plot_bulk_disk_zcoord} examines the origin of the bulk
force on the disc in Fig.~\ref{fig:plot_bulkforce_zcoord}.  The
left-hand panels of this figure show results for a disc centred on
the origin, while the right-hand panels show what happens when the
disc is offset by $\Delta z = 5/64$, which yields a large bulk force.
Top panels show accelerations, plotting the $z$-component of each
particle's error against its $z$ position.  The complex structure
visible in these scatter-plots confirms that adjacent particles have
correlated errors.  If the disc is symmetrically placed with respect
to the tree, as on the left, the error distribution is also symmetric;
particles below (above) the disc's midplane feel spurious downward
(upward) forces.  Conversely, if the disc is displaced, as on right,
the distribution is still correlated, but no longer symmetric.

The bottom panels of Fig.~\ref{fig:plot_bulk_disk_zcoord} show how the
bulk forces accumulate if the second sum in eq.~(\ref{eq:bulk_force})
is ordered by the $z$-coordinate of each particle.  On the left, the
errors due to particles below the midplane are almost perfectly
canceled by those above.  As a result, the bulk force is very small;
it would vanish entirely if not for low-level asymmetries due to
sampling the density field with a finite number of particles.  This
error cancellation is spoiled if the disc is shifted along the $z$
axis; as the right-hand panel shows, the result is a substantial bulk
force in the positive $z$ direction.

\begin{figure}
  \centering
  \includegraphics[width=\linewidth]{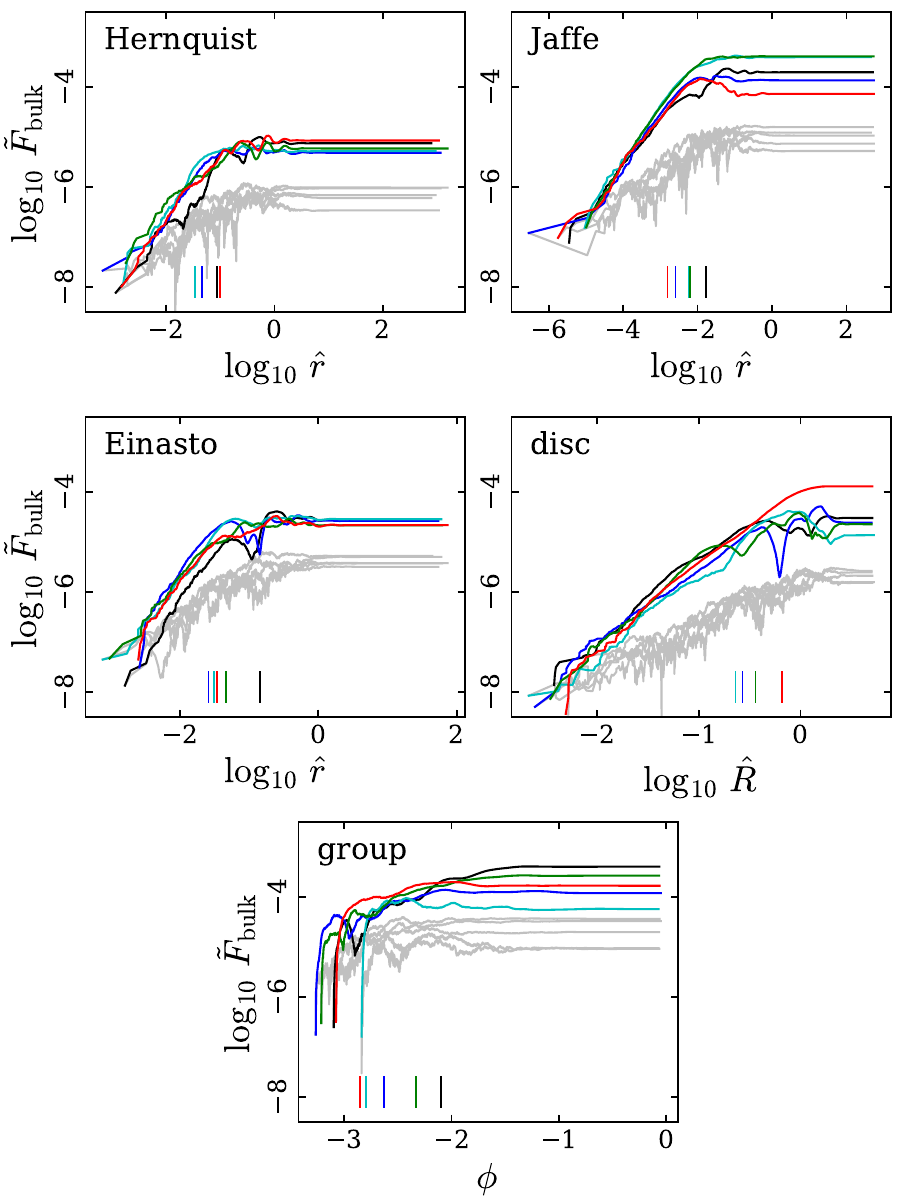}
  \caption{Accumulation of bulk forces in all five mass models.
    Forces in each panel are summed by the variable used for the
    horizontal axis; colored curves are actual results, while light
    grey curves show random walks with the same stepsizes.  All
    realizations have $N = 2^{18}$ particles; forces are calculated
    with $\theta = 0.8$.  Short vertical marks at the bottom of each
    panel show where the bulk force reaches half of its final value.}
  \label{fig:plot_force_cumulative}
\end{figure}

Although the disc model provides perhaps the clearest example of bulk
forces, it is not unique.  Fig.~\ref{fig:plot_force_cumulative} shows
how individual force errors accumulate to yield bulk forces for all
five mass models.  In each panel, the sum in eq.~(\ref{eq:bulk_force})
is ordered by the quantity shown on the horizontal axis: distance
$\hat{r}$ from the model centre for the three spherical models,
cylindrical radius $\hat{R}$ for the disc models, and direct-sum
potential $\phi$ for the group models.  The five colored curves show
actual bulk forces, while the light grey curves were generated by
replacing each $\Delta \tilde{\mathbf{a}}_{p}$ in
eq.~(\ref{eq:bulk_force}) by a randomly-oriented vector of the same
magnitude.  Thus, the grey curves show how bulk forces would
accumulate \textit{if force errors were uncorrelated}.  In every case,
the actual bulk forces are roughly an order of magnitude greater than
their uncorrelated counterparts, demonstrating the ubiquity of error
correlations and their consequences for violations of Newton's third
law.

Fig.~\ref{fig:plot_force_cumulative} illustrates a few other points as
well.  In each panel, short vertical marks indicate the radius (or
potential) where the bulk force reaches half its final value.  For the
three spherical models, these radii are up to two orders of magnitude
\textit{smaller} than their respective half-mass radii; thus, bulk
forces are dominated by a minority of particles with
larger-than-average force errors (see Fig.~\ref{fig:scat_err_rad}).
In effect, these fictitious forces tug the centres of systems around,
dragging the rest in due course.  It is also striking that the bulk
forces on Jaffe model are $\sim 30$ times larger than on Hernquist
models, even though both have similar relative acceleration errors.
Moreover, while all five members of the Hernquist and Einasto
ensembles have similar bulk forces, the Jaffe ensemble displays a much
larger range; evidently these bulk force are quite sensitive to the
placement of the central $\rho \propto r^{-2}$ cusp relative to the
tree.

The disc models also exhibit large variations within their ensemble.
This is not surprising; Fig.~\ref{fig:plot_bulk_disk_zcoord} already
shows how the bulk force on a disc depends on its position.  In
addition, bulk forces on discs are sensitive to orientation.  The disc
shown in red provides a striking example; as Fig.~\ref{fig:hist_err}
shows, this disc yields \textit{better} accelerations than other the
members of its ensemble do, yet here it generates a bulk force nearly
an order of magnitude \textit{larger}!  This occurs because the close
alignment of this disc with the $x$--$z$ plane produces acceleration
errors which add coherently.  In contrast, the other four disc
realizations all have larger misalignments; in every case, their
acceleration errors are larger, but these errors are less coherent,
resulting in smaller bulk forces.  Moreover, these forces are not
necessarily perpendicular to the disc plane; indeed, the disc shown in
cyan, which has the largest acceleration errors, generates a bulk
force nearly parallel to the disc.

Unlike the other four examples, the group models are not spherically
or rotationally symmetric.  This virtually guarantees that they will
generate large bulk forces, since symmetric error patterns like that
seen in the top-left panel of Fig.~\ref{fig:plot_bulk_disk_zcoord},
which tend to self-cancel, almost never occur.  It's noteworthy that
these forces are substantially larger than those obtained for the
Einasto models making up these groups.  The median potential for this
ensemble is $\phi_\mathrm{med} \simeq -1.63$; just as in the spherical
models, a minority of particles, lying deep within the potential well,
account for most of the bulk forces.  Finally, the heterogeneous
nature of the group models also ensures that there are significant
variations between realizations.

\begin{figure}
  \centering
  \includegraphics[width=0.5\linewidth]{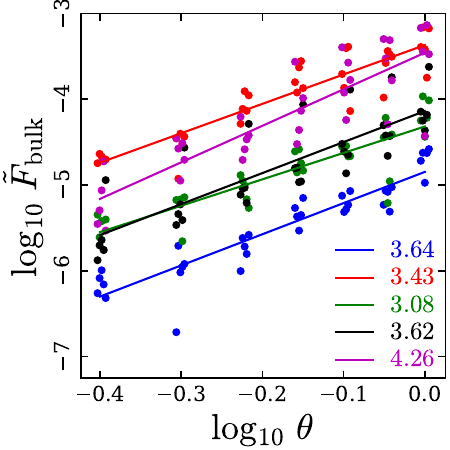}
  \caption{Bulk forces vs.~opening angle.  Mass models are realized
    with $N = 2^{18}$ particles.  Individual realizations are plotted
    as dots, color-coded as in Fig.~\ref{fig:scat_err_theta}; solid
    lines are least-square fits with slopes as shown.}
  \label{fig:scat_force_theta}
\end{figure}

Fig.~\ref{fig:scat_force_theta} plots bulk forces vs.~opening angle
for various mass models. This plot indicates that the
results above, obtained with $\theta = 0.8$, generalize to other
opening angles as well.  For the four intrinsically symmetric models,
bulk forces scale as $\tilde{F}_\mathrm{bulk} \propto \theta^{3.4 \pm
  0.2}$, similar to the scaling of $\delta \tilde{a}_\mathrm{rms}$
seen in \S~\ref{sec:errors_vs_opening_angle}.  This may indicate that
the level of error correlation is not very sensitive to the value of
$\theta$.  On the other hand, the group models converge like
$\tilde{F}_\mathrm{bulk} \propto \theta^{4.3}$, which is somewhat
faster than the slope of $\sim 3.5$ they yield for $\delta
\tilde{a}_\mathrm{rms}$.

\subsection{Bulk Torques}
\label{sec:bulk_torques}

Newtonian mechanics tells us that the total torque on an isolated
system vanishes: $\mathbf{T}_\mathrm{bulk} = \sum m_{p} \mathbf{r}_{p}
\times \mathbf{a}_{p} = 0$.  The third law is necessary but not
sufficient for this to be true; besides being equal and opposite,
forces must point in the right \textit{directions}.  For example,
consider a binary system consisting of two spherical, well-separated
galaxies.  The net forces on the two galaxies may be equal and
opposite, but if they are not accelerating \textit{directly} toward
each other, the bulk torque on the system will not vanish.  Thus,
particle-mesh and tree algorithms which enforce Newton's third law can
still violate angular momentum conservation
\citep{H1987,M2017,K+2021}.

In the presence of bulk forces, some care is needed to define bulk
torques.  Imagine a system, whose centre-of-mass is offset from the
origin of the simulation CS by $\mathbf{r}_\mathrm{cm}$, subject to a
treecode-induced bulk force $\tilde{\mathbf{F}}_\mathrm{bulk}$.  The
cross-product of these two vectors has the characteristics of a
torque, but it does not tell us anything about the rate at which the
system's \textit{internal} angular momentum changes.  For that, we
need to look at the bulk torque calculated in the system's
centre-of-mass frame:
\begin{equation}
  \tilde{\mathbf{T}}_\mathrm{bulk} \equiv
  \sum_{p} m_{p} (\mathbf{r}_{p} - \mathbf{r}_\mathrm{cm}) \times
    \tilde{\mathbf{a}}_{p} =
  \sum_{p} m_{p} \overline{\mathbf{r}}_{p} \times
  \Delta \tilde{\mathbf{a}}_{p} \, ,
  \label{eq:bulk_torque}
\end{equation}
where $\overline{\mathbf{r}}_{p}$ is the position relative to the
centre of mass.

Once again, discs provide an intuitive way to see how bulk torques can
arise.  As Fig.~\ref{fig:plot_bulk_disk_zcoord} shows, a disc which
lies in the $x$--$y$ plane but is slightly offset in the $z$ direction
feels a bulk force pushing it further away from the origin.  Consider
now a disc which is not offset but rather slightly tilted; the side
above the $x$--$y$ plane feels an upward force, while the side below
feels a downward force.  In other words, the disc feels a torque which
acts to increase its initial tilt.

\begin{figure}
  \centering
  \includegraphics[width=\linewidth]{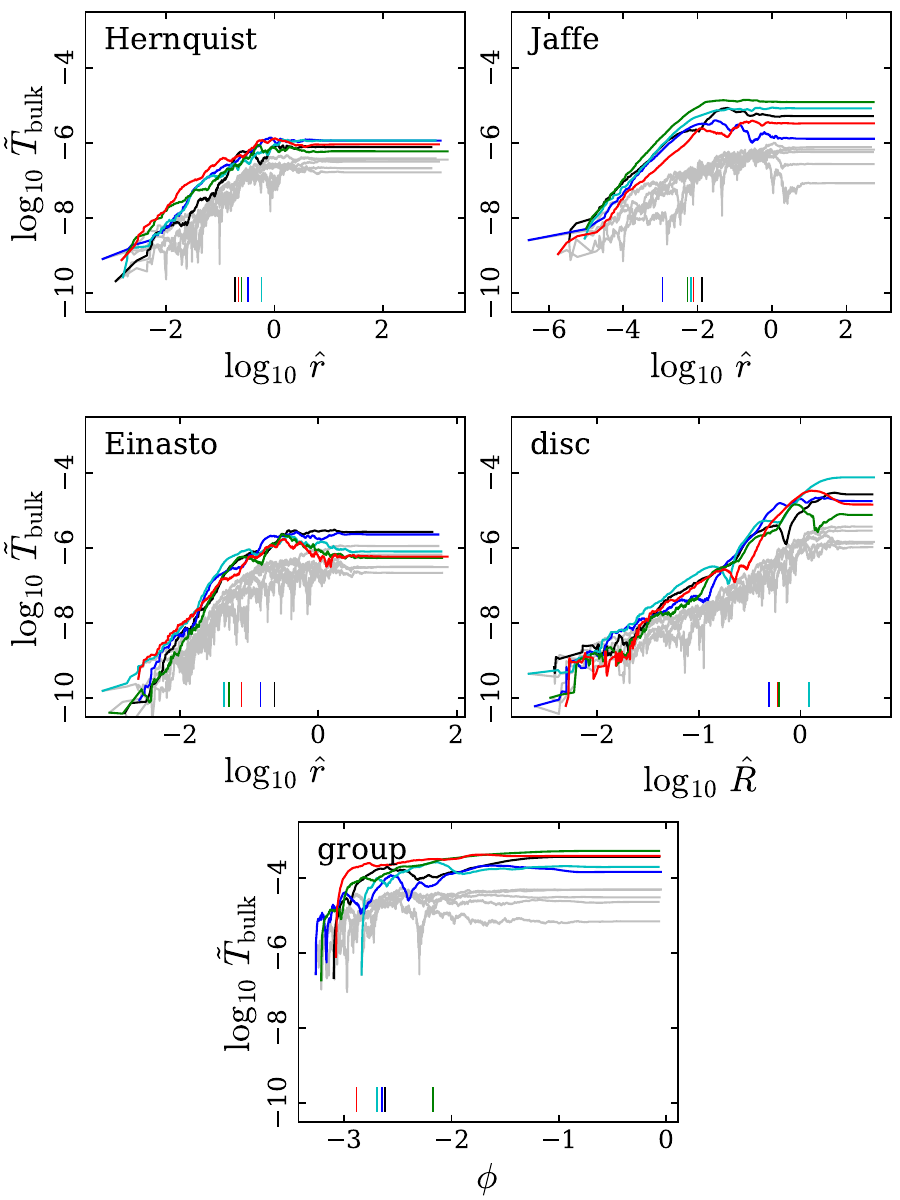}
  \caption{Accumulation of bulk torques in all five mass models.
    Compare with Fig.~\ref{fig:plot_force_cumulative}, which also
    describes plotting conventions.}
  \label{fig:plot_torque_cumulative}
\end{figure}

Fig.~\ref{fig:plot_torque_cumulative} shows how errors accumulate to
yield bulk torques for all five models.  These torques are evaluated
in each realization's centre-of-mass coordinate system.  The
conventions used in this plot are the same used in
Fig.~\ref{fig:plot_force_cumulative}; in particular, colored curves
represent actual torques for each realization, while light grey curves
show random-walk equivalents.

All three spherical models yield bulk torques of fairly similar
magnitude.  The Jaffe model is somewhat of an outlier, with median
torques $\sim 5$ times larger than the other two, and a somewhat
larger scatter between realizations.  In most cases, bulk torques
arise at larger radii than bulk forces, presumably because larger
radii provide longer lever arms.  Finally, actual torques, especially
for the Hernquist and Einasto models, are typically only a few times
larger than those expected in the absence of correlations.  Torques on
spherical models presumably arise from Monte-Carlo
sampling of their mass distributions.  If so, they should scale with
particle number as $T_\mathrm{bulk} \propto N^{-1/2}$.  Actual torques
for the three spherical models are roughly consistent with this
prediction; fitting $\log \, T_\mathrm{bulk}$ to $\log \, N$ yields
power-law indices of $-0.56$, $-0.68$, and $-0.46$ for the Hernquist,
Jaffe, and Einasto models, respectively, with no systematic dependence
on $\theta$.

The disc models experience substantially larger bulk torques than any
of the spherical models do.  As the discussion above implies,
orientations appear to be the main source of the rather sizable
variation among disc realizations, although offsets also contribute.
Actual torques are an order of magnitude larger than uncorrelated
errors predict, with much of this difference arising at scales larger
than $\log_{10} \, \hat{R} > -1$, which is approximately the disc's
scale-height.  Bulk torques on discs appear to be independent of $N$,
which makes sense since irregularities due to Monte-Carlo are not
necessary.

By virtue, again, of their \textit{intrinsic} asymmetries, the group
models yield the largest bulk torques.  Correlations play a
significant role in generating these torques.  Half of the total
torque is exerted on particles deep within the multiple potential
wells of these groups.  Again, these torques show no systematic
dependence on particle number.

\begin{figure}
  \centering
  \includegraphics[width=0.5\linewidth]{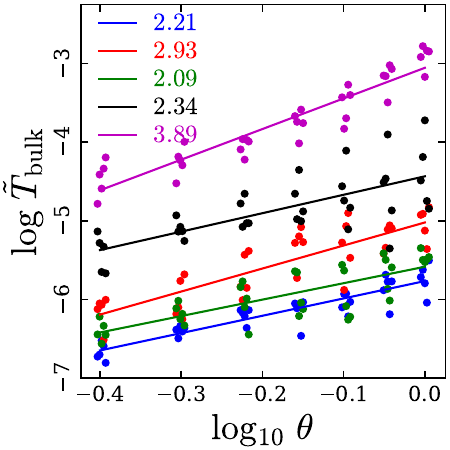}
  \caption{Bulk torques vs.~opening angle.  Compare with
    Fig.~\ref{fig:scat_force_theta}.}
  \label{fig:scat_torque_theta}
\end{figure}

Fig.~\ref{fig:scat_torque_theta} plots bulk torques against opening
angle.  Torques on the spherical models are numerically small, but
decline fairly slowly, roughly as $T_\mathrm{bulk} \propto \theta^{2.0
  \pm 0.1}$.  Disc torques are larger and follow a similar
trend with opening angle.  On the other hand, the torques on the group
models decline  more rapidly, roughly as $\theta^{4.0}$.

\subsection{Computing Time}
\label{sec:computing_time}

\begin{figure}
  \centering
  \includegraphics[width=0.5\linewidth]{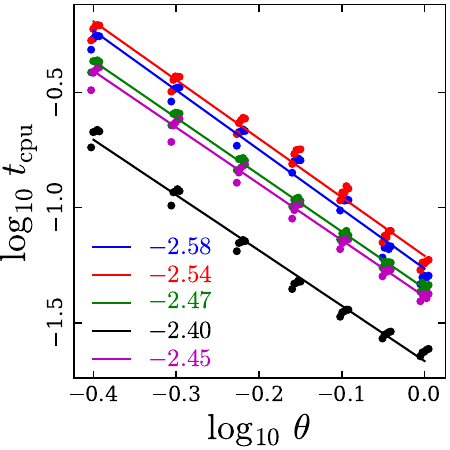}
  \caption{Computing time in minutes vs.~opening angle $\theta$ for
    models color-coded as in Fig.~\ref{fig:scat_err_theta}.  Each
    configuration has $N = 2^{18}$ particles.  Quadrapole terms are
    included.  The thin lines are least-squares fits, with slopes
    shown in the lower left.}
  \label{fig:scat_CPU_theta}
\end{figure}

Fig.~\ref{fig:scat_CPU_theta} shows how computing time\footnote{CPU
times were measured using an Intel~i7 processor with a baseline clock
rate of of $3.40 \, \mathrm{GHz}$.} depends on the opening angle.  All
five models yield rough power-law relationships approximating
$t_\mathrm{cpu} \propto \theta^{-2.5}$.  CPU time for a given $\theta$
is model-dependent; the disc takes roughly a third of the time
required by the Hernquist and Jaffe models.  The latter are slower, in
part because -- even with a taper at radius $b = 100$ -- they include
a few bodies with very large radii. To accommodate these outliers, a
large root cell is required, and it must be subdivided quite deeply to
fully resolve the body distribution.  This makes tree descent more
costly.

\begin{figure}
  \centering
  \includegraphics[width=0.5\linewidth]{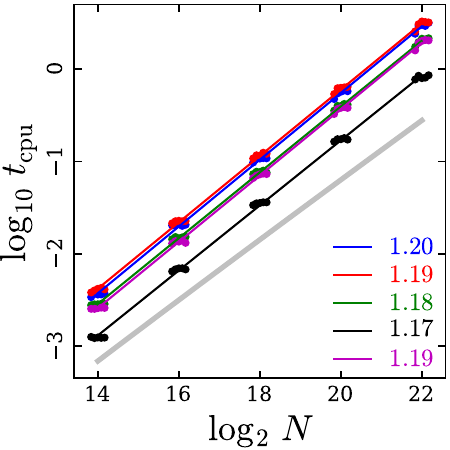}
  \caption{Computing time in minutes vs.~number of particles for
    different models.  Forces were calculated as in
    Fig.~\ref{fig:scat_err_rad}.  These timings are for $\theta=0.8$
    and include quadrapole terms.  Thin lines are least-squares fits,
    with slopes for each shown in the lower right. Plotted $\log N$
    values are randomly offset by small amounts to reduce crowding.
    The light grey curve shows a $N \log(N)$ relationship.}
  \label{fig:scat_CPU_Nbody}
\end{figure}

The power-laws in Figs.~\ref{fig:scat_err_theta}
and~\ref{fig:scat_CPU_theta} can be combined to obtain a relationship
between computing time and accuracy.  For example, eliminating
$\theta$ from $\delta a_\mathrm{rms} \propto \theta^{3.5}$ and
$t_\mathrm{cpu} \propto \theta^{-2.5}$ yields $\delta a_\mathrm{rms}
\propto t_\mathrm{cpu}^{-1.4}$.  This is confirmed by plotting $\delta
a_\mathrm{rms}$ against $t_\mathrm{cpu}$ (see Supplemental
Fig.~3). Remarkably, much of the
model-to-model variation in Figs.~\ref{fig:scat_err_theta}
and~\ref{fig:scat_CPU_theta} cancels out, leaving a near-universal
relationship between computing time and accuracy.

\begin{figure*}
  \centering
  \includegraphics[width=\linewidth]{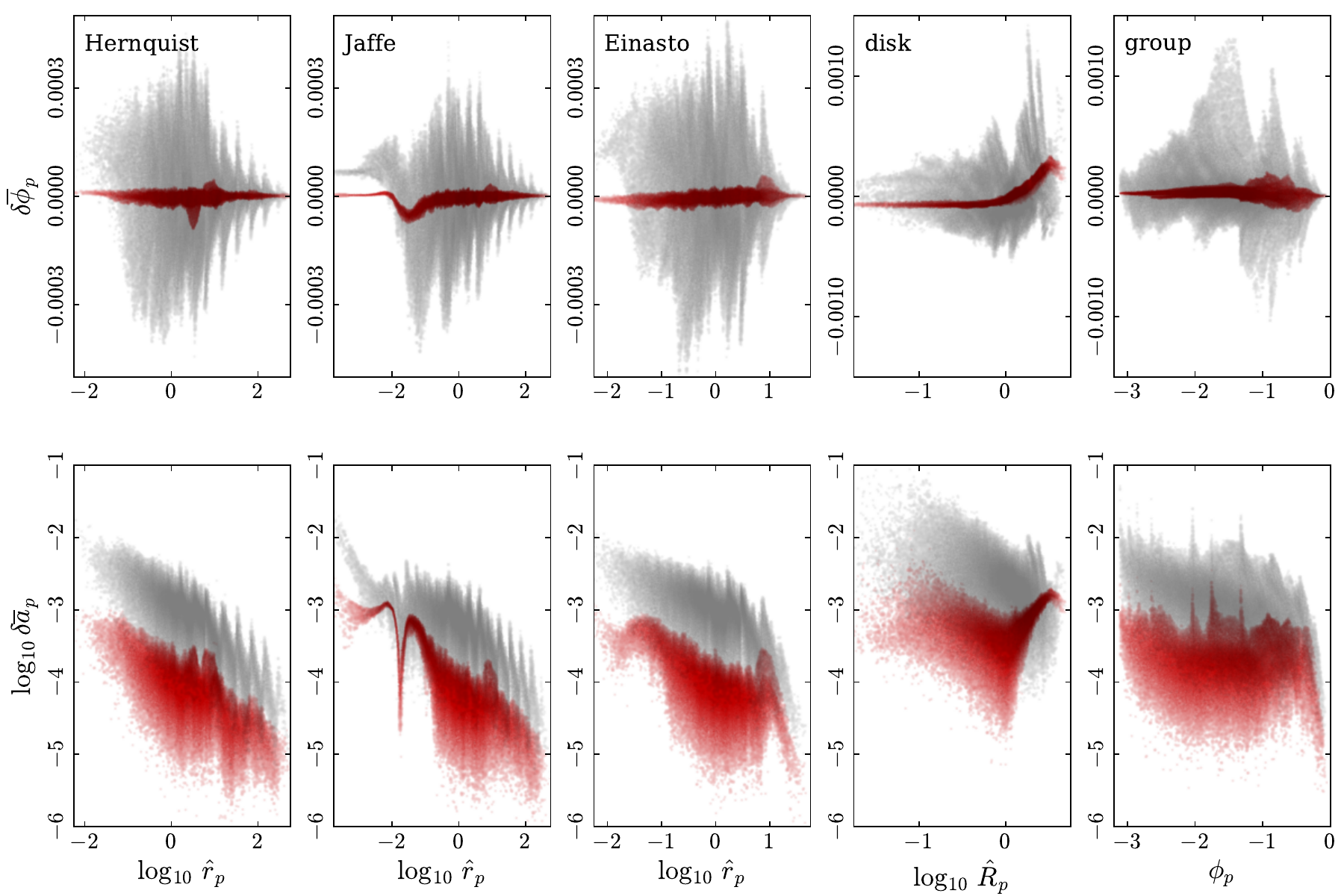}
  \caption{Scatter-plots of force calculation errors versus radius
    or potential.  Layout follows Fig.~\ref{fig:scat_err_rad}.
    Results of averaging over $N_\mathrm{avg} = 256$ trees are shown
    in red, while non-averaged results are shown in grey.}
  \label{fig:scat_err_rad_avg}
\end{figure*}

Finally, Fig.~\ref{fig:scat_CPU_Nbody} examines the relationship
between computing time and number of particles for $\log_2 N = 14, 16,
\dots, 22$.  Over this range of $N$, all five models follow power laws
with slopes which are weakly dependent on $\theta$; for $\theta = 1.0$
the slope is $1.13 \pm 0.02$ (one-sigma), while for $\theta = 0.4$ it
is $1.25 \pm 0.03$.  It makes sense that the slope increases as
$\theta$ decreases; in the limit $\theta \to 0$, the tree-code behaves
like a direct-sum code, implying a slope of $2$.  In the figure, a
light grey curve shows a $t_\mathrm{cpu} \propto N \log(N)$
relationship; this curve has an average slope of $1.08$.  Thus, for
the values of $N$ tested here, the computing time grows
\textit{slightly} faster than $N \log(N)$.

% \clerpage \newpage

\section{TREE AVERAGING}
\label{sec:tree_averaging}

As \S~\ref{sec:static_tests} shows, the forces returned by a single
hierarchical calculation include errors which are peculiar to the
specific tree CS employed.  These errors will partly cancel if results
from multiple, independent trees are averaged together.  Let
($\mathsf{R}_{k}$, $\mathit{S}_{k}$, $\mathbf{T}_{k}$), where $k = 1,
\dots, N_\mathrm{avg}$, be a series of tree CSs chosen at random.  Let
$\tilde{\phi}_{p,k}$ and $\tilde{\mathbf{a}}_{p,k}$ be the potential
and acceleration of particle $p$ computed using tree CS $k$.  Then the
`tree-averaged' potential and acceleration of $p$ are
\begin{equation}
  \overline{\phi}_{p} = \frac{1}{N_\mathrm{avg}}
    \sum_{k=1}^{N_\mathrm{avg}} \tilde{\phi}_{p,k} \, ,
  \qquad
  \overline{\mathbf{a}}_{p} = \frac{1}{N_\mathrm{avg}}
    \sum_{k=1}^{N_\mathrm{avg}} \tilde{\mathbf{a}}_{p,k} \, .
\end{equation}
As the number of independent tree CSs grows, these averages should
better approximate the true potential $\phi_{p}$ and acceleration
$\mathbf{a}_{p}$.

Fig.~\ref{fig:scat_err_rad_avg} shows that averaging can significantly
reduce tree-code errors.  Here the red points show the results of
averaging over $N_\mathrm{avg} = 256$ trees, each employing a
different rotation, scaling, and translation\footnote{For these tests,
I set the maximum translation distance $T_\mathrm{max} = 4$ length
units -- the distance between adjacent sweet spots for the Hernquist
models (\S~\ref{sec:error_distributions}).  This does not seem to be a
critical choice; virtually idential results are obtained for all
models with $T_\mathrm{max} = 1$.}.  The black points, copied directly
from Fig.~\ref{fig:scat_err_rad}, are not averaged.  In every case,
the averaged potentials are much more accurate than their
counterparts.  Likewise, the acceleration errors, plotted on a
logarithmic scale, are about an order of magnitude smaller.  Averaging
has largely erased the oct-tree signatures exhibited by the black
points. For the spherical models, the regular barbs seen in the
$\delta \tilde{\phi}_{p}$ distributions have vanished, and the
sawtooth patterns in the $\delta \tilde{a}_{p}$ distributions are
smoothed out.  The offset in the Jaffe model's central potential,
noted in \S~\ref{sec:error_distributions}, has vanished as a result of
averaging over position.  For the disc and group models, the chaotic
patterns in $\delta \tilde{\phi}_{p}$ are largely suppressed.

All the same, averaging over tree CSs has not \textit{entirely}
eliminated errors in potential and acceleration.  The Jaffe model
provides a clear example, with conspicuous features visible in both
the $\delta \overline{\phi}_{p}$ and $\delta \overline{a}_{p}$
distributions near $\log_{10} \hat{r}_{p} \simeq -1.7$.  The same
pattern of acceleration errors appears to be replicated in the
Hernquist and Einasto models, although less prominently as these
models are sparsely populated at such small radii.  Briefly, the
inward acceleration is \textit{slightly} over-estimated, by at most
$\sim 0.08$~per cent (assuming $\theta = 0.8$), for radii $\log_{10}
\hat{r} \gtrsim -1.7$, and under-estimated, by a comparable amount, at
smaller radii.  The radial scale of these features is comparable to
the softening length $\epsilon = 0.01$, suggesting a link to residual
errors in the softened quadrapole term
(eq.~\ref{eq:quadrapole_potential}).

The disc model offers another kind of example, with positive $\delta
\overline{\phi}_{p}$ within $\log_{10} \hat{R}_{p} \lesssim 0$, and
negative elsewhere.  This is \textit{not} related to softening;
rather, the sign-change occurs near the half-mass radius ($\log_{10}
\hat{R}_\mathrm{half} \simeq 0.07$).  A connection to the physical
scale of the model is bolstered by the acceleration errors (lower
panel), which are suppressed by roughly an order of magnitude for
$\log_{10} \hat{R}_{p} \lesssim 0$, but approach the non-averaged
errors at larger radii.

\begin{figure}
  \centering
  \includegraphics[width=0.75\linewidth]{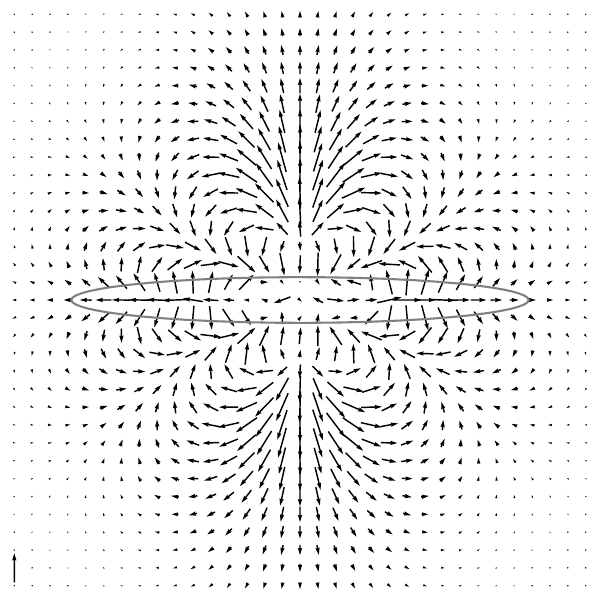}
  \caption{Residual acceleration errors for a Gaussian disc, viewed
    edge-on.  Forces computed with $\theta = 0.8$ were averaged over
    $N_\mathrm{avg} = 256$ CSs.  The oval curve, which represents the
    disc, extends $\pm 4$ and $\pm 0.4$ length units along the $x$ and
    $z$ axes, respectively.  For scale, the vertical arrow at the
    lower left shows an acceleration error $|\Delta
    \overline{\mathbf{a}}| = 0.0003$.}
  \label{fig:quiv_accerr_disk}
\end{figure}

Residual acceleration errors $\Delta \overline{\mathbf{a}} =
\overline{\mathbf{a}} - \mathbf{a}$ generated by the disc model are
mapped in Fig.~\ref{fig:quiv_accerr_disk}.  The global pattern seen
here is independent of $\theta$; only the amplitude changes.  It is
likewise invariant with respect to rotation about the $z$-axis and
reflection across the $xy$-plane.  The octapole component dominates,
with significant higher-order contributions.  Small irregularities
(e.g., within the central few grid-points) arise from (1)
sampling the density field with $N = 2^{18}$ particles, and (2)
averaging over $N_\mathrm{avg} = 256$ tree CSs; tests with different
CS sequences and different realizations indicate that sampling
effects dominate the irregularities.

In Fig.~\ref{fig:quiv_accerr_disk}, one can easily find closed paths
around which line integrals are non-zero; thus $\Delta
\overline{\mathbf{a}}$ is not the gradient of any function.  Since the
true acceleration is the gradient $\mathbf{a} = \nabla \phi_\epsilon$,
where $\phi_\epsilon$ is the softened N-body potential, it follows
that $\overline{\mathbf{a}}$ is not a perfectly conservative force
field.

These error patterns become evident for any reasonably large value of
$N_\mathrm{avg}$.  They arise because oct-tree gravity calculations
make \textit{systematic} errors which are not tied to the choice of
coordinate system.  Rather, such errors are generic to the whole
\textit{approximation scheme}.  A particle at the outskirts of a thin
disc interacts with many cells which enclose substantial chunks of the
disc, and therefore have large octapole-plus moments.  This is true
regardless of the CS used to construct the tree, so its consequences
will be manifest even after averaging over multiple CSs.

\begin{figure*}
  \centering
  \includegraphics[width=\linewidth]{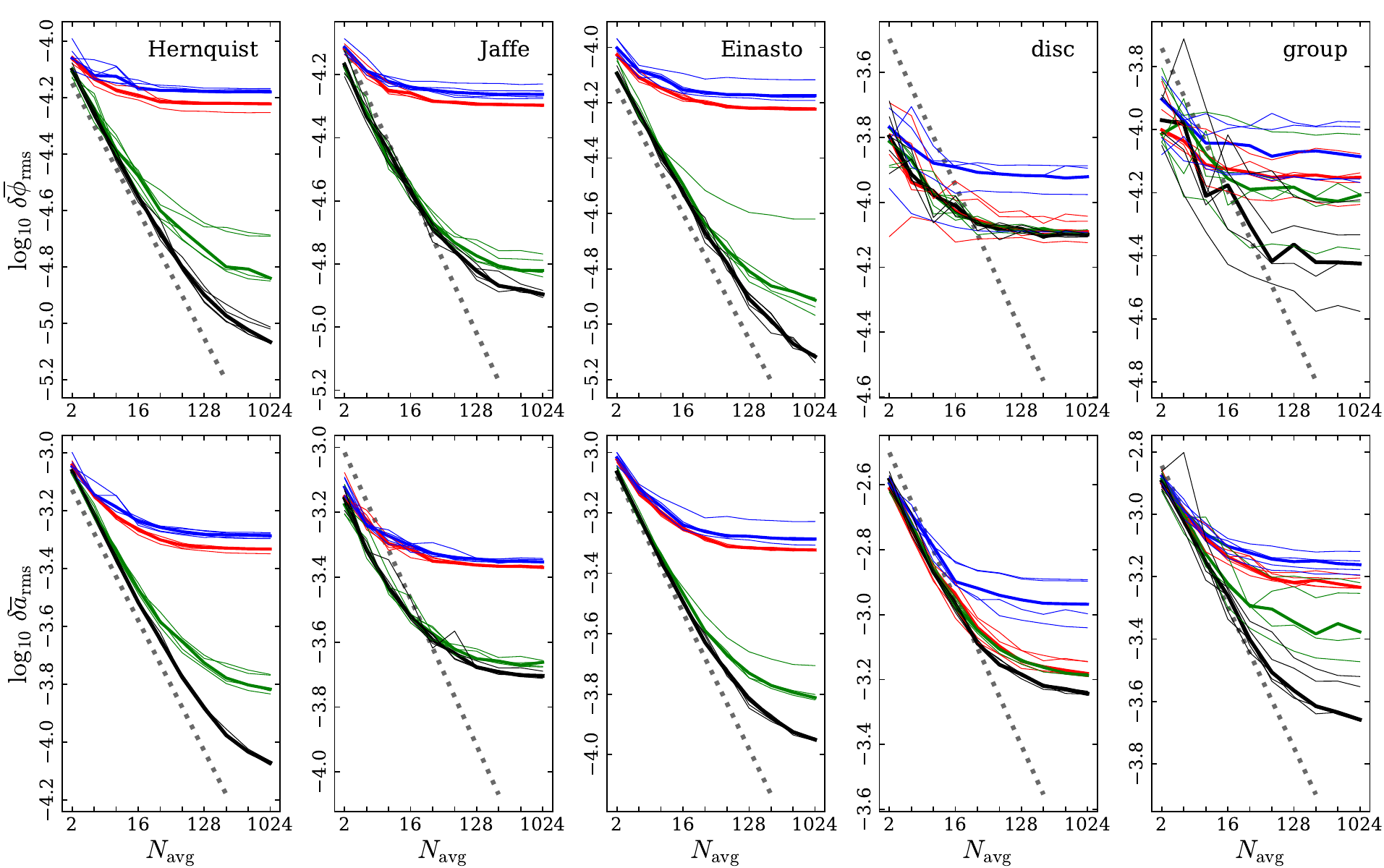}
  \caption{RMS errors in tree-averaged potentials (top) and
    accelerations (bottom) vs.~number of averages, using $N = 2^{18}$
    particles and $\theta = 0.8$.  In each panel, black curves show
    results for all three transformations together; red, green, and
    blue curves show results for translation, rotation, and scaling,
    respectively; a grey dotted line shows the
    scaling $\propto N_\mathrm{avg}^{-1/2}$ expected from the central
    limit theorem.  Thin lines represent individual realizations,
    while heavier lines show medians.}
  \label{fig:plot_err_navg_all}
\end{figure*}

To see how errors in the averaged potential and acceleration depend on
the number of averages, I define aggregated RMS errors $\delta
\overline{\phi}_\mathrm{rms}$ and $\delta \overline{a}_\mathrm{rms}$
by analogy with eq.~(\ref{eq:def_agg_error}).
Fig.~\ref{fig:plot_err_navg_all} presents log-log plots of these
errors versus $N_\mathrm{avg}$.  A different seed was used to generate
the set of random transformations ($\mathsf{R}_{k}$, $\mathit{S}_{k}$,
$\mathbf{T}_{k}$) for each realization and choice of $N_\mathrm{avg}$.
Thus, while the same five realizations were tested for multiple values
of $N_\mathrm{avg}$, the transformations used are independent.  In
each panel, thin lines show individual realizations, a heavy black
line shows the median relationship between $N_\mathrm{avg}$ and
$\delta \overline{\phi}_\mathrm{rms}$ or $\delta
\overline{a}_\mathrm{rms}$ obtained using rotation, translation, and
scaling together, while a grey dotted line shows a power-law $\propto
N_\mathrm{avg}^{-1/2}$.  This is the relationship the central limit
theorem would predict in the absence of systematic errors.

For modest values of $N_\mathrm{avg}$, the averaged potential and
acceleration errors for the spherical models do in fact scale roughly
as $N_\mathrm{avg}^{-1/2}$.  However, as $N_\mathrm{avg}$ grows, the
errors `peel away' from this ideal trend and seem to be leveling off.
As a conceptual model, the error for a given particle may be split
into two parts: a \textit{random} part which depends on the tree CS,
and a \textit{systematic} part which is independent of the tree CS.
Averaging suppresses the random part by a factor of
$N_\mathrm{avg}^{-1/2}$, but has no effect on the systematic part.  In
the Hernquist and Einasto models, random errors are much larger than
systematic ones, so averaging provides significant improvements before
the systematic errors dominate.  Systematic acceleration errors appear
larger for the Jaffe models; as a result, the $\delta
\overline{a}_\mathrm{rms}$ curves deviate from $N_\mathrm{avg}^{-1/2}$
earlier.

If systematic errors are comparable to random ones then tree averaging
will provide limited benefits.  This is presumably the case for the
disc and group potentials: median errors decrease by factors of $\sim
2$ and $\sim 2.5$ respectively as $N_\mathrm{avg}$ increases, compared
to the order-of-magnitude reductions the spherical models enjoy.
Moreover, the group models exhibit especially large variations between
realizations, implying that the layout of the mass distribution
strongly influences the level of systematic error.  However, the
corresponding acceleration errors (bottom panel) exhibit better
convergence; this is encouraging as accurate accelerations are more
critical for simulations than accurate potentials.

The results described so far employ all three tree transformations:
rotation, scaling, and translation.  How do these transformations
contribute individually?  The red, green, and blue curves in
Fig.~\ref{fig:plot_err_navg_all} address this question by plotting
$\delta \overline{\phi}_\mathrm{rms}$ and $\delta
\overline{a}_\mathrm{rms}$ as functions of $N_\mathrm{avg}$ for each
transformation separately.  For the spherical models, the upshot is
that rotation (green) is by far the most effective, accounting in each
case for a large part of the overall improvement, while translation
(red) and scaling (blue) provide smaller and roughly comparable
benefits.  One of the Einasto realizations benefits less from rotation
than do the other four; it has a smaller offset ($r_\mathrm{off}
\simeq 1.23$, compared to a mean of $3.18$ for the rest), so random
rotations don't displace its centre as much.  Clearly, rotation alone
will confer little benefit for spherical models situated near the
origin.

A different pattern emerges for the disc models.  Rotation and
translation appear to be equally effective, and each in isolation is
almost as good as all three transformations together.  By itself,
scaling offers somewhat greater benefits than it does for the
spherical models, with noticeable variations in acceleration errors
among the five realizations.

For the group models, rotation again provides the largest gain in
accuracy, especially for accelerations.  However, this benefit varies
significantly from one realization to the next.  In contrast,
translation and scaling yield modest improvements, of comparable
magnitude, with less variation among realizations.

\begin{figure}
  \centering
  \includegraphics[width=\linewidth]{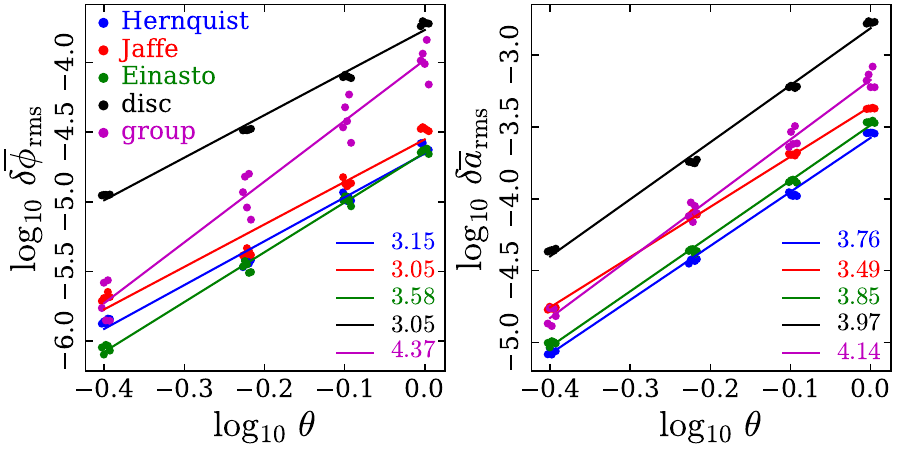}
  \caption{Tree-averaged RMS errors versus log opening angle.
    $N_\mathrm{avg} = 256$ averages were used for each
    calculation. Compare with right panels of
    Fig.~\ref{fig:scat_err_theta}.}
  \label{fig:scat_err_theta_avg}
\end{figure}

Do the advantages of averaging persist for other choices of opening
angle?  Fig.~\ref{fig:scat_err_theta_avg} shows how RMS errors in
averaged potential and acceleration depend on $\theta$.  Comparison
with Fig.~\ref{fig:scat_err_theta} confirms that averaging and opening
angle work well together; errors averaged over $N_\mathrm{avg} = 256$
trees decrease as least as fast as un-averaged errors as $\theta$ is
reduced.  Forces for Hernquist and Einasto models become a factor of
$\sim 10$ more accurate; the other models improve by factors of $\sim
6$ or more.

\begin{figure*}
  \centering
  \includegraphics[width=\linewidth]{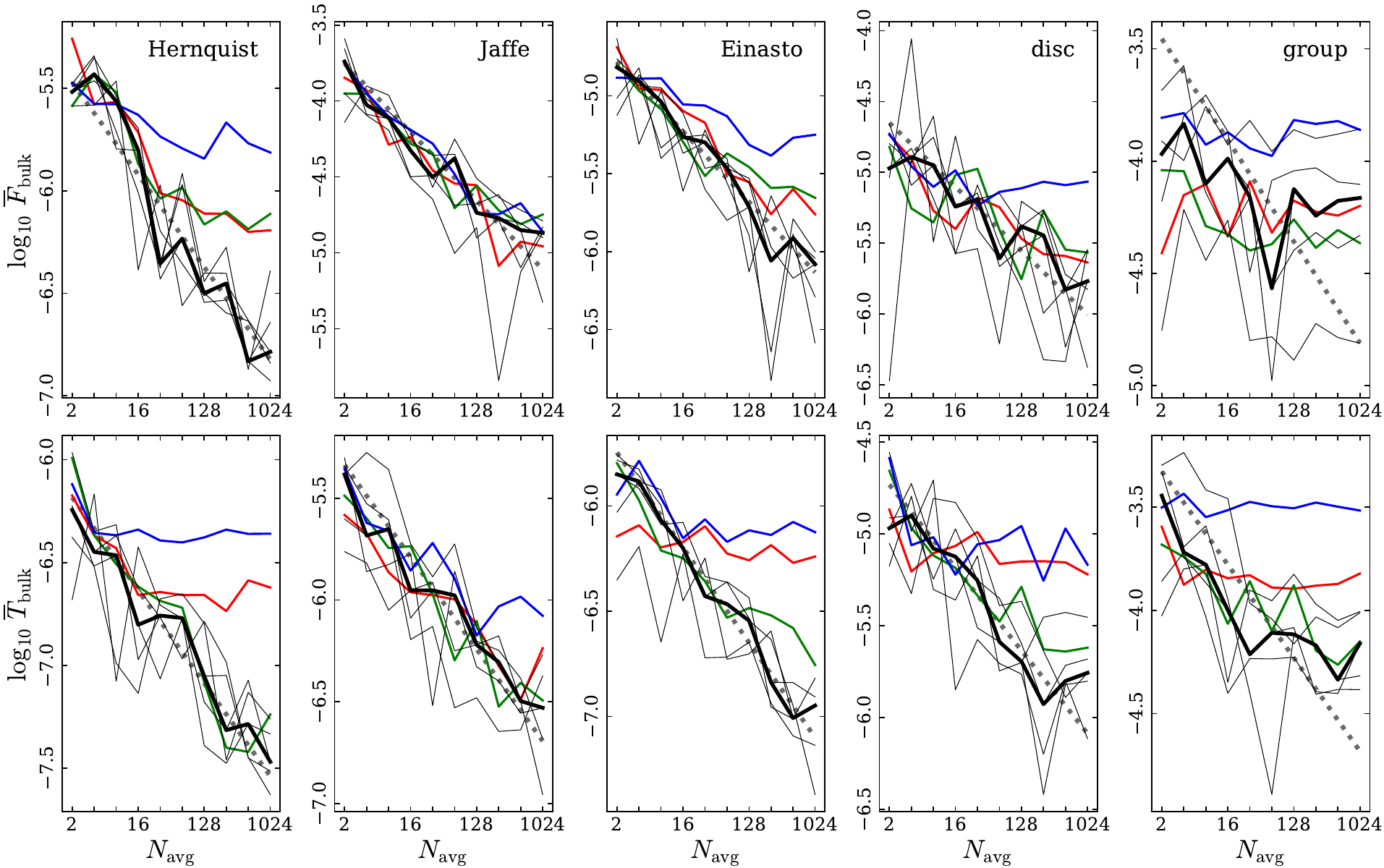}
  \caption{Tree-averaged bulk forces (top) and torques (bottom)
    vs.~number of averages.  Plotting conventions follow
    Fig.~\ref{fig:plot_err_navg_all}, however, individual realizations
    (thin lines) are only shown for all three transformations together
    to reduce crowding and confusion.}
  \label{fig:plot_bulk_navg_all}
\end{figure*}

Fig.~\ref{fig:plot_bulk_navg_all} show how bulk forces and torques
behave as the number of random tree CSs increases.  Since bulk forces
and torques are residuals left after near-cancellation between $N$
approximate accelerations, the results tend to be somewhat noisy.
Forces and torques obtained using all three transformations are shown
in black: thick lines are medians, while thin lines track individual
realizations.  Grey dotted lines show slopes of $-1/2$.  Red, green,
and blue lines show medians obtained using each of the three
transformations separately; individual realizations are not shown to
limit clutter.

Compared to the others, bulk forces on groups are relatively
\textit{insensitive} to averaging.  For $N_\mathrm{avg} \ge 128$, the
force on each of the five group realizations converges toward a
nonzero value specific to that particular system's layout.  Averaging
is somewhat more effective at reducing torques, but here too the
results appear to level off as $N_\mathrm{avg}$ increases.  It seems
that bulk forces and torques on groups \textit{do not vanish}, no
matter how may tree CSs are averaged over.

\begin{figure}
  \centering
  \includegraphics[width=0.75\linewidth]{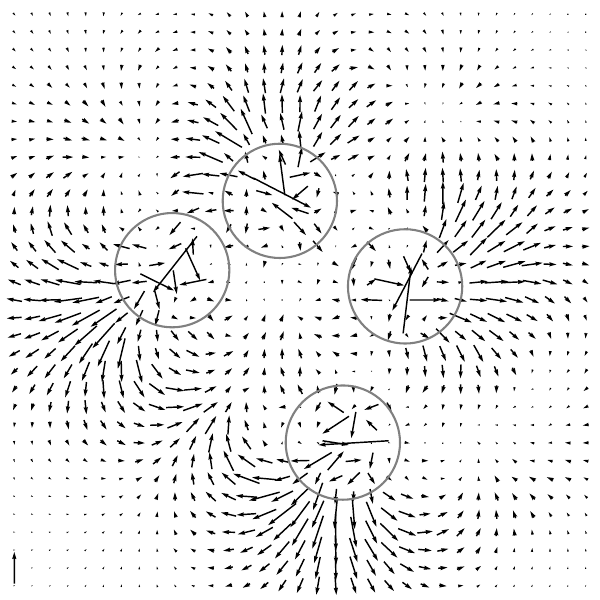}
  \caption{Residual acceleration errors for a planar group.  All four
    Einasto models lie in the same plane, viewed face-on.  Forces
    computed with $\theta = 0.8$ were averaged over $N_\mathrm{avg} =
    256$ CSs.  Unit circles show positions and half-mass radii of
    group members.  The vertical arrow at the lower left shows an
    acceleration error $|\Delta \overline{\mathbf{a}}| = 0.0003$.}
  \label{fig:quiv_accerr_group}
\end{figure}

At first glance, this may seem surprising.  Averaging \textit{should}
restore rotational and translational invariance; how can these group
models still be subject to bulk forces?  The answer, ultimately, goes
back to \citeauthor{H1987}'s (\citeyear{H1987}) observation that a
particle and a distant cell don't exert equal and opposite forces;
again, this is true regardless of the tree CS, so averaging doesn't
mitigate it.  A symmetric mass distribution induces a symmetric
pattern of averaged force errors, so bulk forces effectively cancel
out.  But the group models are too asymmetric for this to happen (see
Fig.~\ref{fig:quiv_accerr_group}); instead, residual forces can add
coherently.  The bulk torques on the groups presumably arise in much
the same way.

Bulk forces and torques on the spherical and disc models track
$\overline{F}_\mathrm{bulk} \propto N_\mathrm{avg}^{-1/2}$ and
$\overline{F}_\mathrm{bulk} \propto N_\mathrm{avg}^{-1/2}$
\textit{remarkably} well.  These models have highly symmetric mass
distributions, so residual forces cancel almost perfectly.  Of course,
any realization constructed with a finite number of particles has
irregularities which weakly violate these symmetries.  Thus at some
point, forces and torques on these realizations should peel away from
the $N_\mathrm{avg}^{-1/2}$ scaling law, and eventually level, off
much as the group models do.  Bulk forces on the Jaffe realizations do
appear to be leveling off for $N_\mathrm{avg} \ge 512$.  The discs may
also be reaching the limits of their $N_\mathrm{avg}^{-1/2}$ scaling,
again because of fluctuations due to finite $N$.  Realizations with
more particles should track $N_\mathrm{avg}^{-1/2}$ scaling even
better than the examples shown here; unfortunately, testing this is
expensive!

In passing, it's striking that bulk forces and torques on the
spherical and disc models follow $N_\mathrm{avg}^{-1/2}$ scaling well
beyond the point where the acceleration errors for these models
(Fig.~\ref{fig:plot_err_navg_all}) are clearly leveling off.  This
illustrates the point that bulk forces and torques encode information
above and beyond RMS acceleration errors.

Fig.~\ref{fig:plot_bulk_navg_all} also uses red, green, and blue lines
to show how bulk forces and torques respond to averaging using each
transformation separately: red for translation, green for rotation,
and blue for scaling.  For the spherical and disc models, rotation and
translation appear about equally effective at mitigating bulk forces.
Naively, one might expect that bulk forces should respond more to
translation than to rotation.  A likely explanation of the actual
results is that by potentially displacing a realization's centre to
any point on a sphere of radius $r_\mathrm{off}$, rotation provides
benefits similar to translation.  Conversely, rotation is usually
superior to translation at mitigating torques; the exception is the
Jaffe model, which seems to benefit about equally from both
transformations.

\begin{figure}
  \centering
  \includegraphics[width=\linewidth]{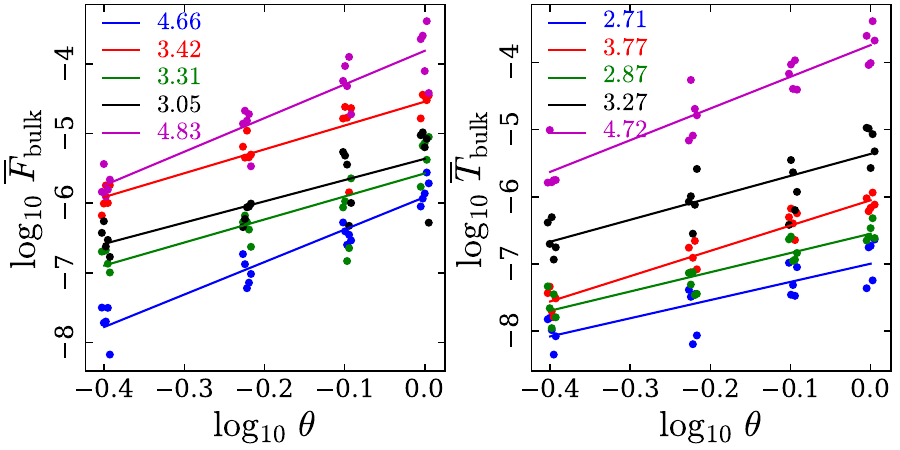}
  \caption{Bulk forces and torques vs.~opening angle, obtained by
    averaging over $N_\mathrm{avg} = 256$ tree CSs.  Compare with
    Figs.~\ref{fig:scat_force_theta} and \ref{fig:scat_torque_theta}.}
  \label{fig:scat_bulk_theta_avg}
\end{figure}

Fig.~\ref{fig:scat_bulk_theta_avg} shows how bulk forces and torques,
averaged over $N_\mathrm{avg} = 256$ tree CSs, depend on opening
angle.  Comparison with Figs.~\ref{fig:scat_force_theta} and
\ref{fig:scat_torque_theta} again confirms that averaging and opening
angle work well together.  Bulk forces and torques on spherical and
disc models are reduced by an order of magnitude.  Due to their
intrinsic asymmetry, groups benefit somewhat less, but their bulk
forces and torques fall off rapidly as $\theta$ decreases.

% \clerpage \newpage

\section{DISCUSSION}
\label{sec:discussion}

Tree averaging is an example of a general strategy in numerical
calculation: \textit{don't reuse a single approximation} if
alternativres exist.  An oct-tree provides a convenient set of
approximations for evaluating gravitational forces, but if the tree is
always built using the same CS, the same approximations will be used
again and again.  Randomizing over a plurality of tree CSs breaks this
repetitive pattern, delivering more accurate results.  For highly
symmetric systems, accelerations computed by averaging can be an order
of magnitude more accurate than those obtained from a single CS.  The
gains for asymmetric systems (i.e., groups) are somewhat smaller but
still significant.  In addition, averaging makes tree-codes more
accurately invariant under translations and rotations.

Yet averaging is not a universal panacea: once the accidental errors
arising from a specific choice of tree CS have been suppressed,
a different category of error, intrinsic to \textit{any} hierarchical
force calculation scheme, comes to light.  In particular, averaging
does not address the fact that highly flattened or stratified mass
distributions generate cells with large octapole-plus moments, or
prevent an array of such cells, all sampling the same mass
distribution, from producing correlated errors.  Nor does averaging
resolve the asymmetry between a cell $c$ acting on a particle $p$ and
a particle $p$ acting on the particles $p^\prime \in c$ \citep{H1987};
thus averaged force calculations can still yield nonzero bulk forces
and torques unless interactions are explicitly symmetrized
\citep{D2000,M2017,K+2021}.

The relationship between momentum conservation and galilean invariance
is straightforward if forces are conservative: Noether's theorem shows
that conservation laws are consequences of invariance.  But
Fig.~\ref{fig:quiv_accerr_disk} demonstrates that forces averaged over
multiple tree CSs are \textit{not conservative}.  Thus a tree-code may
approach invariance but not conserve momentum; the unequal forces a
particle and a cell exert on each other are still in effect.  For
example, a configuration consisting of a Gaussian sphere and a thin
disc, separated by $2$ length units along the minor axis of the
latter, reliably generates a strong bulk force toward the sphere;
after averaging, this force is independent of the system's position or
orientation, demonstrating translational and rotational invariance.

Lagrangian tree-codes using trees constructed via a nearest-neighbor
scheme \citep{BBCP1990} might seem to have an advantage here; the tree
is already invariant under translation, rotation, and scaling.
However, this built-in invariance still permits violations of Newton's
third law.  Moreover, it's not clear how to incorporate other benefits
of averaging; each particle configuration generates a \textit{unique}
tree, in contrast to the plethora of trees an Eulerian codes can
generate via an affine transformation (eq.~\ref{eq:tree_coords}).  Is
it worth introducing randomness into a nearest-neighbor tree to
decorrelate errors?

The idea behind averaging is very general, and it may be useful for a
wide variety of codes.  Any tree, adaptive refinement, or mesh code
which implements vacuum boundary conditions (BC), typically employed
for simulations of isolated systems (including encounters), could use
rotation and translation with little difficulty; scaling is
problematic for codes with mesh-based resolution.  The advantages may
depend on the type of code; e.g., Cartesian mesh codes may benefit
from rotation, since they don't conserve angular momentum accurately
\citep{H1987}.

Tree-codes implementing periodic BCs via Ewald summation can't use
random rotation; the rotated field density field does not preserve
periodic BCs. Scaling is precluded, since the root cell must be
identified with the simulation volume.  But translation is relatively
straightforward; \cite{SPZR2021} describe a code which does exactly
that, optionally in conjunction with a momentum-conserving force
calculation algorithm \citep{D2000}.  On the other hand, periodic-BC
codes which use a particle-mesh algorithm to handle long-range forces
and a tree-code to `fill in' short-range forces \citep{X1995,
  Bagla2002, S2005} are \textit{effectively} using the latter to solve
a vacuum-BC problem, and may be able to deploy all three
transformations.

The averaging procedure described here chooses tree CSs at random.  Is
this the best strategy?  Random sampling spans the entire space of
possibilities, but not efficiently.  Iterating over a well-chosen
sequence of tree CSs might be better, much as panel integration is
typically more accurate than Monte-Carlo integration. However,
choosing the CS sequence seems non-trivial; it may depend on both the
system to be simulated and the specific objectives of the simulation.
On the other hand, randomness is easy to implement and minimizes a
priori assumptions.

Likewise, the opening criteria in \S~\ref{sec:opening_criteria}
minimize a priori assumptions, basically aiming to constrain the
relative error of each cell.  Conversely, GADGET \citep{SYW2001}
prioritizes the accuracy of the cells which exert the largest forces.
This seems reasonable, and is very likely to reduce the broad spread
of acceleration errors shown in the lower right panel of
Fig.~\ref{fig:plot_gravwalk_3D_hernq}, but it builds in some
assumptions about simulation objectives.  As an alternative, it may be
useful to reduce $\theta$ for cells with large octapole-plus moments,
identified either by explicit calculation or by heuristics (e.g.,
standard deviation of mass in descendent nodes).

\section{CONCLUSIONS}
\label{sec:conclusions}

A single oct-tree provides a representation of an $N$-body mass
distribution which can be used to compute approximate gravitational
forces.  Averaging over a \textit{plurality of representations} partly
cancels errors and thereby improves the results of these computations,
while revealing a substratum of more fundamental errors.

The tree-code described in this paper implements averaging. I
characterize the code's performance in detail, aiming to demonstrate
limitations as well as advantages.  \S~\ref{sec:methods} presents the
algorithm; the main innovation is simply to apply a random
angle-preserving affine transformation to particle coordinates before
constructing the tree.  This offers no advantages for a single force
calculation; rather, the advantages accrue over multiple force
calculations.

The systematics of tree-code force calculation are explored in
\S~\ref{sec:static_tests}.  A key point, which emerges early, is that
force-calculation errors are correlated at several levels.  First of
all, the forces that different cells exert on a given particle have
errors which are not statistically independent.  Second, adjacent
particles typically interact with similar sets of cells, and
consequently may have have similar force-calculation errors.

\S~\ref{sec:tree_averaging} examines the results of averaging over
tree CSs.  Averaging suppresses errors which depend on the position
and orientation of the root cell, improving the accuracy of potentials
and accelerations and substantially reducing bulk forces and torques.
On the other hand, errors which are independent of the tree CS are
immune to averaging.  For example, discs give rise to larger force
errors because they are represented by trees containing cells with
large octapole moments; averaging improves disc forces but exposes an
eightfold pattern of errors (Fig.~\ref{fig:quiv_accerr_disk}) which
are likely intrinsic to any tree code using quadrapole but not
octapole moments.

In any case, averaging over multiple tree CSs is an inefficient way to
improve force calculations for \textit{static} systems.  The CPU time
is proportional to the number of averages, so errors scale like
$t_\mathrm{cpu}^{-1/2}$ at best.  In contrast, changing $\theta$
yields errors which scale roughly as $t_\mathrm{cpu}^{-1.4}$.  For
static systems, averaging restores rotational and translational
invariance, but at a very high cost.  Rather, the idea is to implement
averaging in dynamical simulations by using a different tree CS for
each time-step.  However, it's not a given that randomizing tree CSs
\textit{while particles move} will work as well as it does for static
configurations.  The next paper will examine how tree averaging
performs in various dynamic simulations; some preliminary results
appear below.

\subsection{Preliminary Dynamical Tests}
\label{sec:prelim_dynam_tests}

Quantifying the accuracy of an $N$-body simulation is problematic,
since we lack reference solutions for $N > 2$ particles.  However,
simulations which conserve energy, momentum, and angular momentum
within narrow tolerances are generally preferable to those which do
not.

\begin{figure}
  \centering
  \includegraphics[width=\linewidth]{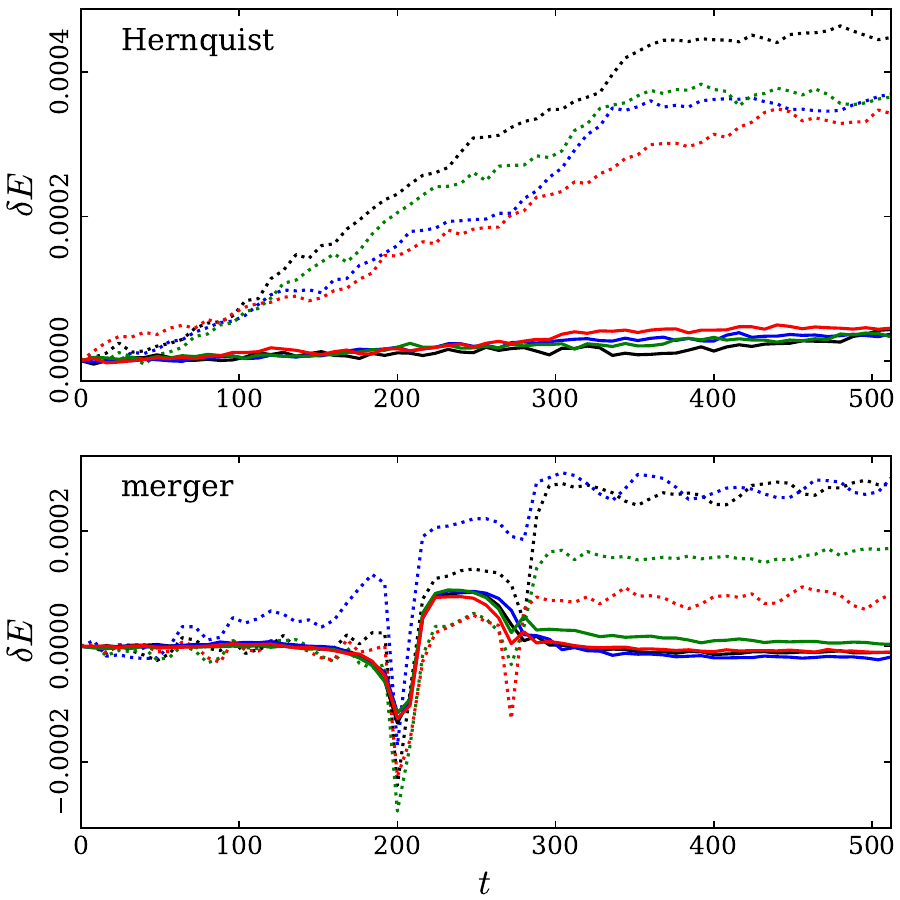}
  \caption{Relative energy conservation errors vs.~simulation
      time for (top) isolated Hernquist models and (bottom) parabolic
      mergers of Hernquist models.  Solid and dotted lines show
      results with and without averaging, respectively; colors simply
      distinguish realizations.}
  \label{fig:plot_deltaE_time_examp}
\end{figure}

Fig.~\ref{fig:plot_deltaE_time_examp} (top panel) shows that averaging
can improve energy conservation by an order of magnitude in
simulations of spherical systems.  Here, energy conservation is gauged
by $\delta E \equiv (E(t) - E(0)) / |E(0)|$, where $E(t)$ is the net
binding energy\footnote{To estimate $E(t)$ more accurately, the
gravitational energy was recalculated from particle snapshots saved
every $8$ time units using $\theta = 0.5$.}.  Eight Hernquist models,
each containing $N = 2^{18}$ particles, were each run for $\sim 15$
half-mass orbital periods, using a time-step $\Delta t = 1/256$, an
opening angle $\theta = 1$, and a softening length $\epsilon = 0.01$.
Solid and dotted lines show results for averaged-tree and fixed-tree
simulations, respectively.

The large differences between the runs with and without averaging
strongly implicates force calculation as the main driver of these
energy conservation errors.  Other tests (not shown) indicate that
errors due to the time-step or numerical precision ($32$-bit
floating-point) set a floor on $|\delta E|$ at least an factor of ten
\textit{below} the best outcomes shown here.  It may be appropriate to
think of these force errors as an external `heat source' driving the
net binding energy $E$ upward toward zero.  It's worth noting that the
errors at each time-step are comparable in all eight simulations; in
the averaged runs, these errors have less effect because they are
decorrelated correlated from one time-step to the next.

(Parenthetically, it's curious that the un-averaged runs seem to level
off as they approach $\delta E \simeq 0.0004$.  Why this happens is
unclear; there's no obvious structural change which could moderate the
effects of force errors toward the end of these runs.  The variation
among these experiments also lacks explanation.  Further examination
of these points is beyond this paper's scope.)

Mergers of equal-mass systems are represented in the bottom panel of
Fig.~\ref{fig:plot_deltaE_time_examp}.  Eight pairs of isotropic
Hernquist models were placed on inbound parabolic orbits reaching
pericentres $r_\mathrm{peri} = 2$ length units apart at
$t_\mathrm{peri} \simeq 200$ time units \citep{SB2024}.  These runs
used the modified treecode (Appendix~\ref{sec:modified_tree_code}),
with $N = 2^{18}$ particles, $\Delta t = 1/256$, $\theta = 0.8$, and
$\epsilon = 0.01$.  Again, averaged and fixed-tree simulations are
represented by solid and dotted lines.

As above, averaging improves energy conservation, but now the picture
is more complicated.  In both the averaged and fixed-tree simulations,
$\delta E$ drops abruptly when the models first encounter each other
($t \simeq 200$), and rebounds as they separate; some of the
fixed-tree simulations seem to repeat this pattern when the systems
merge ($t \simeq 280$).  The sign of this variation suggests that the
net accelerations of the two models are slightly but systematically
underestimated when they are close to each other; as a result, the
models arrive at their first pericentric passage with slightly less
kinetic energy than they should have.

Closer examination sharpens contrasts.  The four averaged runs track
each other \textit{very} closely.  The four runs without averaging,
besides responding more strongly at first encounter, display large
run-to-run variations; in particular, some of these simulations
exhibit sporadic increases in $\delta E$, finally leveling off after
the models merge.  These variations, while comparable in magnitude to
the gradual upward drift in $\delta E$ seen in the top panel, probably
have a different origin.  In particular, the large run-to-run
differences indicate that $\delta E$ is sensitive to the precise
trajectories of the models with respect to the fixed tree structure.
Averaging, by restoring translational invariance, mitigates such
effects.

How significant are the conservation errors seen in
Fig.~\ref{fig:plot_deltaE_time_examp}?  One natural scale is the
overall variation in binding energy between statistically independent
realizations, which in relative terms is $\delta E \sim N^{-1/2}
\simeq 0.002$.  This is a factor of $5$ to $10$ larger than any of the
energy errors found here; by this standard, all of these experiments
are fairly accurate.

But energy conservation is not the only standard.  Conservation of
angular momentum is critical in simulations of binary systems
\citep{M2017,K+2021}.  For example, a $1$~per cent change in the
magnitude of the orbital angular momentum of a merging encounter
yields a change of comparable magnitude in the time between first and
second pericentres.  The fixed-tree merger experiments violate orbital
angular momentum conservation at the $0.1$~per cent level, enough to
contribute to run-to-run variations in merger time-scales
\citep{SB2024}.  In contrast, the four merger experiments with
averaging conserve orbital angular momentum to one part in $10^4$ or
better.

\section*{ACKNOWLEDGMENTS}

I thank Jeremy Goodman, Colby Haggerty, Lars Hernquist, Piet Hut, Jun
Makino, Kegio Nitadori, Atsushi Taruya, and Simon White for
helpful discussions, and the referee for a constructive
report. I'm also grateful to Minghui Chen for reviving the
decade-old hardware used for most of these experiments.  Technical
support and advanced computing resources from University of
Hawai\raisebox{\dimexpr\fontcharht\font`A-\height}{\scalebox{0.8}{`}}i
Information Technology Services – Research Cyberinfrastructure, funded
in part by the National Science Foundation CC* awards \# 2201428 and
\# 2232862, are gratefully acknowledged. Publication costs were
  covered by the Institute for Astronomy.

\section*{DATA AVAILABILITY}

An open-source implementation of the simulation code is available at
\textsf{https://github.com/barnes-astro/Treecode2}. Please direct
requests for the test data and analysis code to the author,
\textsf{barnes@hawaii.edu}.

% \clerpage \newpage

% \newpage

\appendix

\section{MODIFIED TREE-CODE}
\label{sec:modified_tree_code}

Even with threading, scanning the tree is expensive.
Algorithm~\ref{code:iterative_tree_scan} visits only $O(\log N)$ nodes
to compute the force on a particle, but these nodes are scattered in
memory and are accessed in random order.  Moreover, the sequence of
operations is hard to predict, thwarting many of the optimizations
attempted by modern compilers.

Physically adjacent particles typically interact with very similar
sets of nodes.  This suggests a straight-forward way to reduce the
number of tree scans \citep{B1990}: construct a list of interactions
which applies everywhere within some small neighborhood, and reuse it
for each particle contained therein.  The `modified' algorithm finds
such neighborhoods by scanning the tree, looking for cells containing
no more than $n_\mathrm{share}$ particles, where $n_\mathrm{share}
\sim 10^2$ is a user-specified parameter.  Each time it finds a
suitable cell $c'$, it lists the particles inside, and evaluates their
mutual interactions.  It then performs a single tree scan to find
nodes outside $c'$ with which all the particles in $c'$ can interact,
storing the relevant data in two arrays of particles and cells.  These
arrays are then used to sum the external field on each particle in
$c'$; everything is in a linear format, so memory access and
instruction pipelining can be efficiently optimized.  The linear
organization of the data is also compatible with multi-threading, GPU
execution, or specialized $N$-body hardware \citep{M1991}.

The version of the algorithm presented here differs slightly from the
earlier version \citep{B1990}.  To decide if particles in $c'$ could
safely interact with cell $c$, the old version needed the minimum
distance between the volumes represented by $c'$ and $c$.  That
distance is easily evaluated in the tree's CS, but harder to compute
in simulation coordinates, making the $\mathsf{MustOpen}$ function
more costly and explicitly dependent on details of tree construction.
Instead, the new code employs an (almost) coordinate-free criterion.
First, a sphere which bounds the particles in $c'$ is identified.
This sphere is centred at $\mathbf{r}_\mathrm{mid}$, which is found by
averaging minimum and maximum particle positions along all three
simulation coordinates; its radius $r_\mathrm{max}$ is the maximum
distance of any particle in $c'$ from $\mathbf{r}_\mathrm{mid}$.
Then, during the tree scan, a cell $c$ at a distance $s =
|\mathbf{r}_\mathrm{mid} - \mathbf{r}_{c}|$ \textit{must} be opened if
\begin{equation}
  s < \ell_{c} / \theta + d_{c} + r_\mathrm{max} \, .
  \label{eq:mod_offset_criterion}
\end{equation}
This condition guarantees that every particle in $c'$ can interact
with $c$ without falling afoul of eq.~(\ref{eq:offset_criterion}).
For a given $\theta$, it typically increases the number of
particle-cell interactions computed by $\sim 20$~per cent, and the
number of particle-particle interactions by a factor of a few.  This
somewhat improves the algorithm's accuracy.  The effect on computing
time will depend on the computing architecture employed; some specific
results for a single-threaded implementation on a conventional CPU are
briefly covered below.

An unintended side-effect of eq.~(\ref{eq:mod_offset_criterion}) is
that \textit{every} cell $c$ within radius $s < r_\mathrm{max}$ of
$\mathbf{r}_\mathrm{mid}$ \textit{must} be opened.  In situations
where $c'$ is embedded in a steep, large-scale density gradient, this
radius can extend beyond $c'$ and enclose an absurdly large number of
particles, which must be included in the external interaction list.
Rather than allocating arrays large enough to deal with this very rare
situation, the modified code aborts the tree walk if more than a fixed
number of external particles and cells are found.  In the event of an
abort, the algorithm simply processes the immediate descendents of
$c'$ instead of $c'$ itself.  Since these cells are smaller than $c'$,
their $r_\mathrm{max}$ radii are guaranteed to enclose fewer external
particles, eventually resolving the problem.  By default, space is
allocated for a total of $20000$ particles and cells, which is
generous in light of typical usage; this allocation may be worth
tuning if tree-walks are frequently aborted, which is more likely for
calculations with smaller opening angles (e.g., $\theta \lesssim
0.4$).

\subsubsection*{Modified tree-code tests}
\label{sec:modified_tree_code_tests}

The modified code avoids recursive tree scans in favor of iteration
over linear arrays, which can be more easily optimized, vectorized, or
parallelized.  One simple gauge of the algorithm's performance is the
average number of particles which share a list.  This average is $\sim
0.28 n_\mathrm{share}$, with little sensitivity to the mass model,
particle number for $N \ge 2^{18}$, or value of $n_\mathrm{share} \in
[16,256]$.

\begin{figure}
  \centering
  \includegraphics[width=\linewidth]{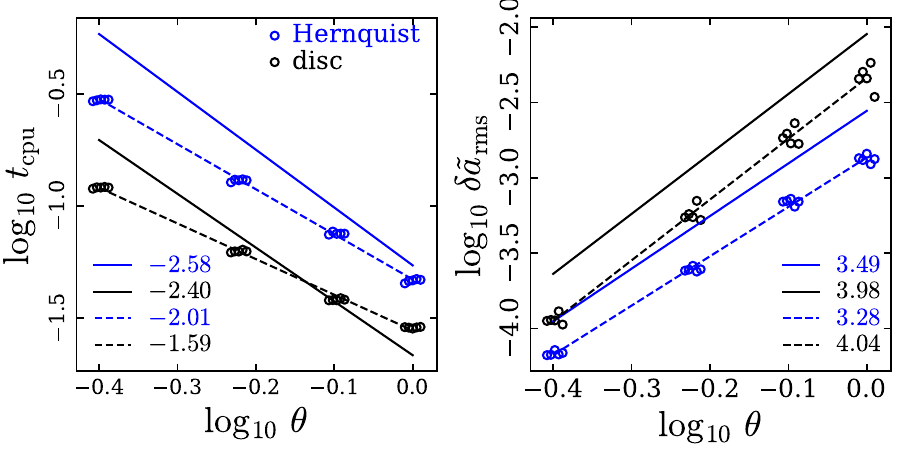}
  \caption{Modified tree-code (dashed lines, open circles) compared to
    standard tree-code (solid lines).  Left: CPU time vs.~opening
    angle.  Right: acceleration error vs.~opening angle.  All tests
    used $N=2^{18}$ particles; the modified code used
    $n_\mathrm{share} = 64$.}
  \label{fig:scat_CPU_acc_theta_mod}
\end{figure}

\begin{figure}
  \centering
  \includegraphics[width=\linewidth]{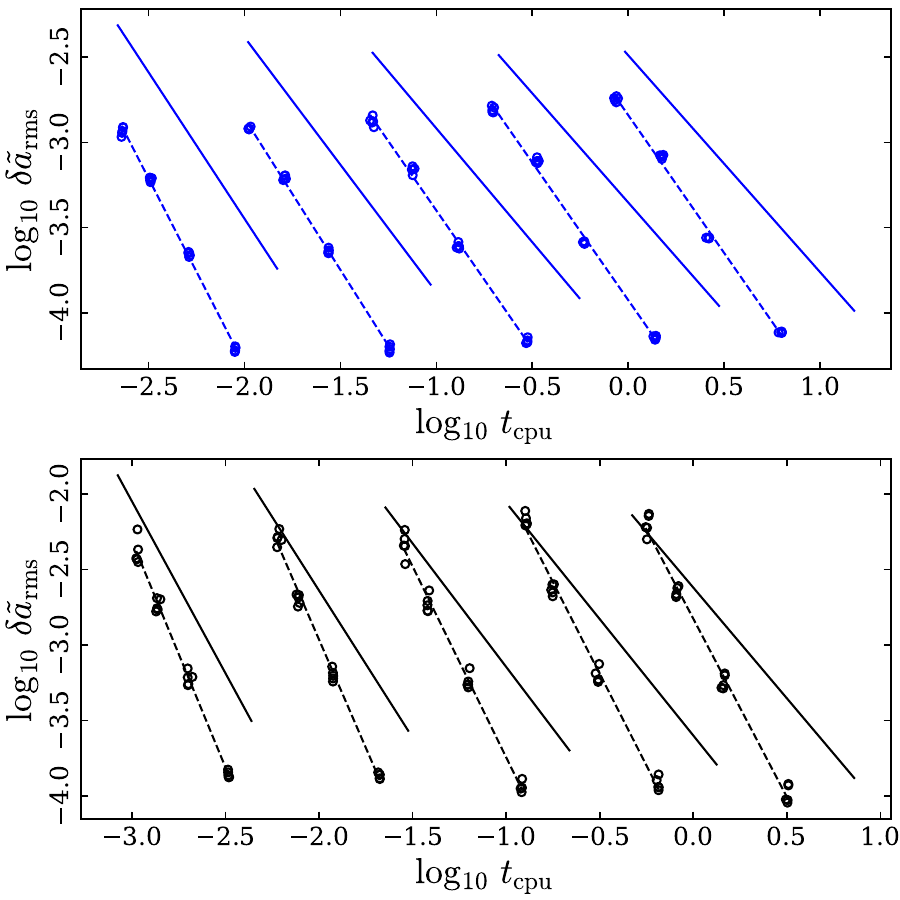}
  \caption{Acceleration error versus computing time for modified (open
    circles, dashed lines) and standard (solid lines) tree-codes.
    Open circles sample the relationship between $t_\mathrm{cpu}$ and
    $\delta \tilde{a}_\mathrm{rms}$ for $\theta = 1.0, 0.8, 0.6, 0.4$;
    lines are fits.  Five values of $N$, progressing from $N = 2^{14}$
    (left) to $N = 2^{22}$ (right), are shown.  Top panel plots
    Hernquist models, bottom panel shows disc models.}
  \label{fig:scat_accrms_CPU_mod}
\end{figure}

\begin{figure*}
  \centering
  \includegraphics[width=\linewidth]{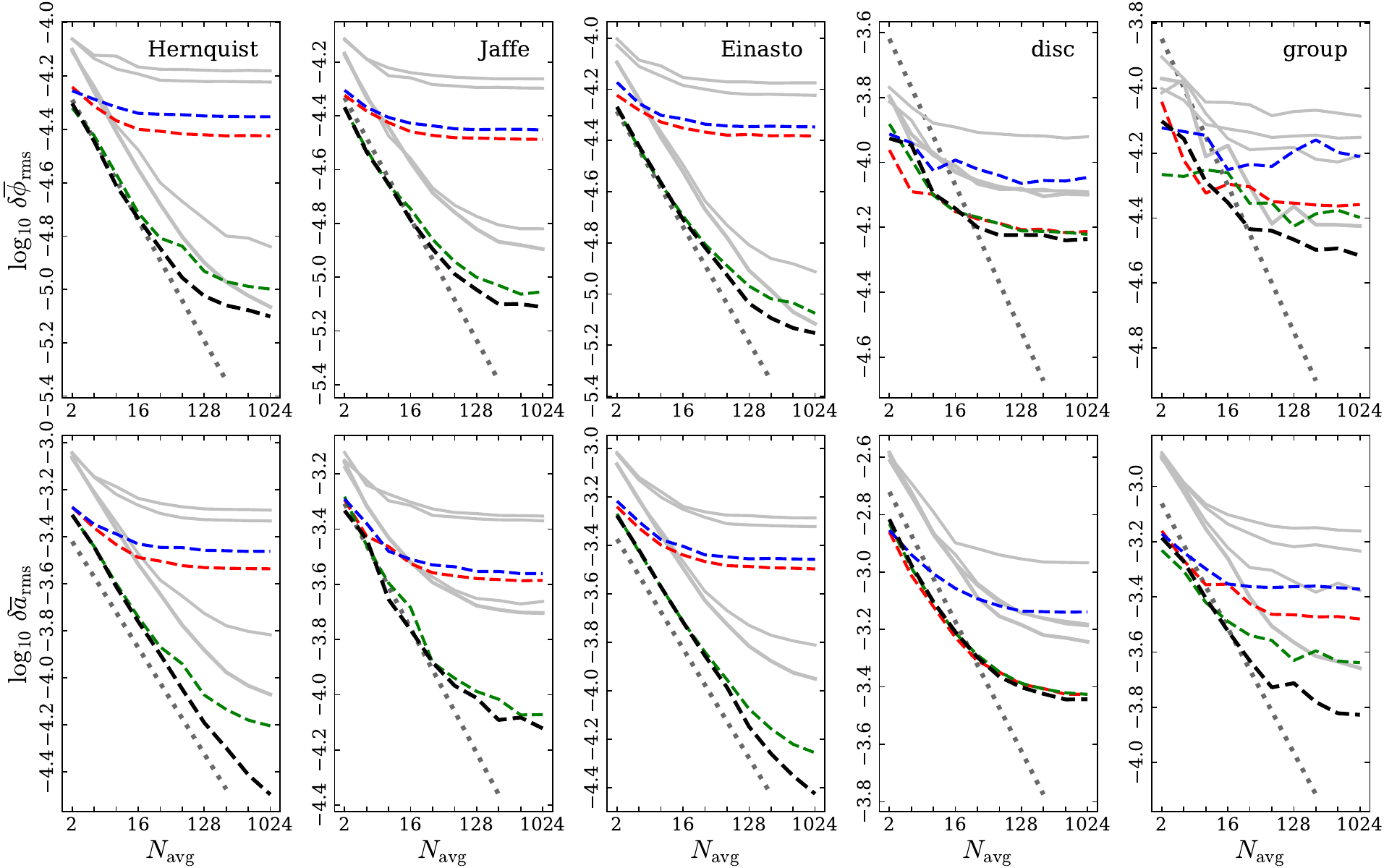}
  \caption{Dashed curves show RMS errors in tree-averaged
      potentials (top) and accelerations (bottom) vs.~number of
      averages, computed using the modified code.  Solid grey curves
      show corresponding results for the standard code from
      Fig.~\ref{fig:plot_err_navg_all}.}
  \label{fig:plot_err_navg_all_mod}
\end{figure*}

Increasing $n_\mathrm{share}$ reduces the number of tree scans, but
\textit{increases} the number of nodes per average interaction list.
At some point, the growing cost of summing over the interactions
exceeds the saving afforded by fewer tree scans.  The total cost per
force calculation has a broad minimum for intermediate values of
$n_\mathrm{share}$.  This optimal value for $n_\mathrm{share}$ depends
on hardware and implementation details; for example, if the force
summation is accelerated (e.g., by multi-threading or special-purpose
hardware), the minimum CPU time will be reached for a larger
$n_\mathrm{share}$.  Preliminary results for a single CPU-only system
indicate $n_\mathrm{share} \simeq 64$ is a good choice, with little
sensitivity on other parameters.

In the left panel of Fig.~\ref{fig:scat_CPU_acc_theta_mod}, I compare
CPU times for the single-thread implementation of the modified code
and the standard version.  The speed-up depends on the mass model and
$\theta$, and to some degree on the value of $N$ as well.  For the
Hernquist model, the modified code is always faster, while for the
disc model, the standard code is faster for $\theta <
\theta_\mathrm{be} \simeq 0.7$, but slower for smaller $\theta$
values.  This break-even point shifts to $\theta_\mathrm{be} \simeq
0.8$ ($0.4$), for $N = 2^{22}$ ($2^{14}$) particles.

Increasing the number of nodes on the interaction list also improves
the accuracy of force calculations.  This is shown on the right in
Fig.~\ref{fig:scat_CPU_acc_theta_mod}, which plots RMS acceleration
errors vs.~$\theta$ for the same ensembles shown on the left.  For any
given $\theta$ value, the modified code is about twice as accurate as
the standard one.  Similar gains are realized for other mass models.
This improvement results largely from better treatment of interactions
on relatively short scales -- a few times the diameter of the set of
particles sharing an interaction list.  As $N$ increases, these
short-range interactions provide a smaller fraction of the total
force, and the gain in accuracy gradually decreases.

Although the modified code was originally developed to take advantage
of vectorized or parallel architectures, it outperforms the standard
code even on scalar CPUs.  This is demonstrated in
Fig.~\ref{fig:scat_accrms_CPU_mod}, which plots RMS force error
vs.~computing time.  Each slanting line shows how these quantities
vary for $\theta$ values within $[0.4, 1.0]$, holding the number of
particles fixed.  Solid and dashed lines represent the standard and
modified code, respectively.  The left-most pair of solid and dashed
lines show results for $N = 2^{14}$ particles; each step to the right
represents a four-fold increase in $N$, up to $N = 2^{22}$ on the
right.  For any given level of accuracy, the modified code is faster,
although the speedup is never much more than a factor of two, and
approaches unity for disc models with large-$N$ and $\theta = 1$.
This modest speed-up arises because the force-summation data is
organized into linear arrays, which allow the compiler to generate
more efficient code.  Larger gains can be expected if force summations
are parallelized or non-CPU hardware can be used.

Finally, is the modified code compatable with tree-averaging?
Fig.~\ref{fig:plot_err_navg_all_mod} plots RMS errors in potential and
acceleration versus number of averages.  This figure's layout
parallels Fig.~\ref{fig:plot_err_navg_all}.  Dashed curves show
modified tree-code results, to be compared with the solid grey curves
presenting median results for the standard code. For any given number
of averages $N_\mathrm{avg}$, the modified-code yields consistently
better results than those produced by the standard code.  In addition,
averaged accelerations for the spherical models track the
$N_\mathrm{avg}^{-1/2}$ trend better than their standard counterparts.

\bsp

\label{lastpage}

\end{document}

\bibliographystyle{plainnat}

\title{Treecode2: The Power of Pluralism\\
Supplemental Figures}
\author{Joshua E. Barnes}
\date{\small \today}
\maketitle

\begin{figure}[h!]
  \centering
  \includegraphics[width=0.75\textwidth]{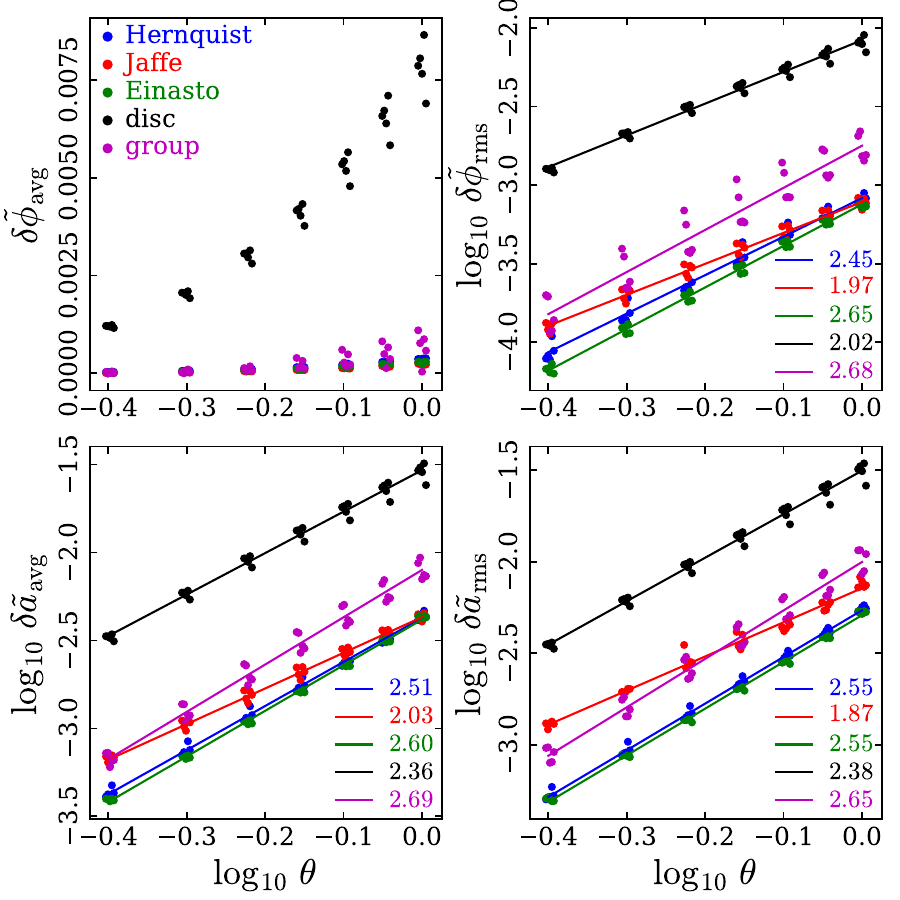}
  \caption{Similar to Fig.~6, but computed
    \textit{without} quadrapole corrections.}
  \label{fig:scat_err_theta_noquad}
\end{figure}

\begin{figure}[h!]
  \centering
  \includegraphics[width=0.5\textwidth]{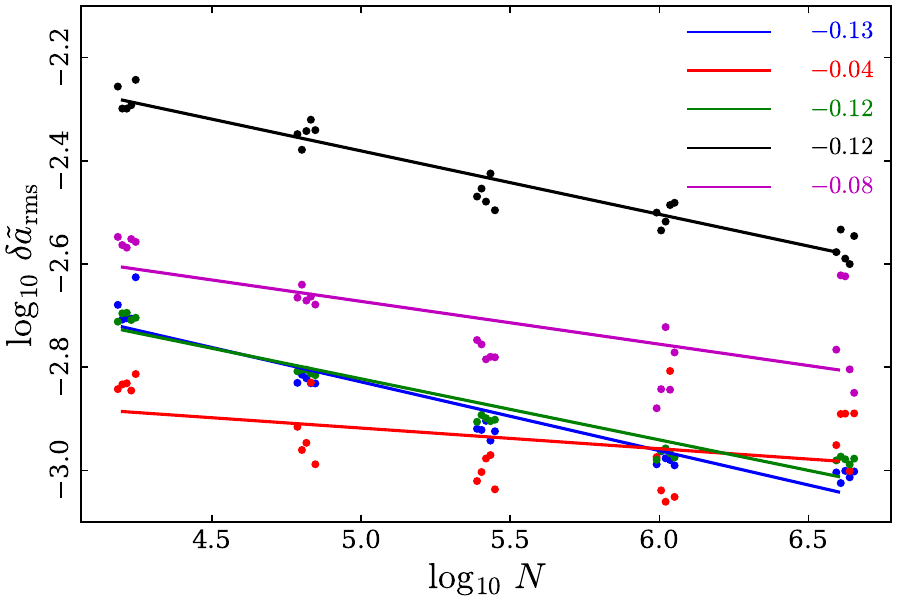}
  \caption{RMS acceleration errors, computed for $\theta = 0.8$,
    vs.~number of particles.  Color indicates mass model as in
    Fig.~6.  Symbols show individual
    configurations, while lines are power-law fits with slopes as
    indicated.}
  \label{fig:scat_accrms_Nbody}
\end{figure}

\begin{figure}[h!]
  \centering
  \includegraphics[width=0.5\textwidth]{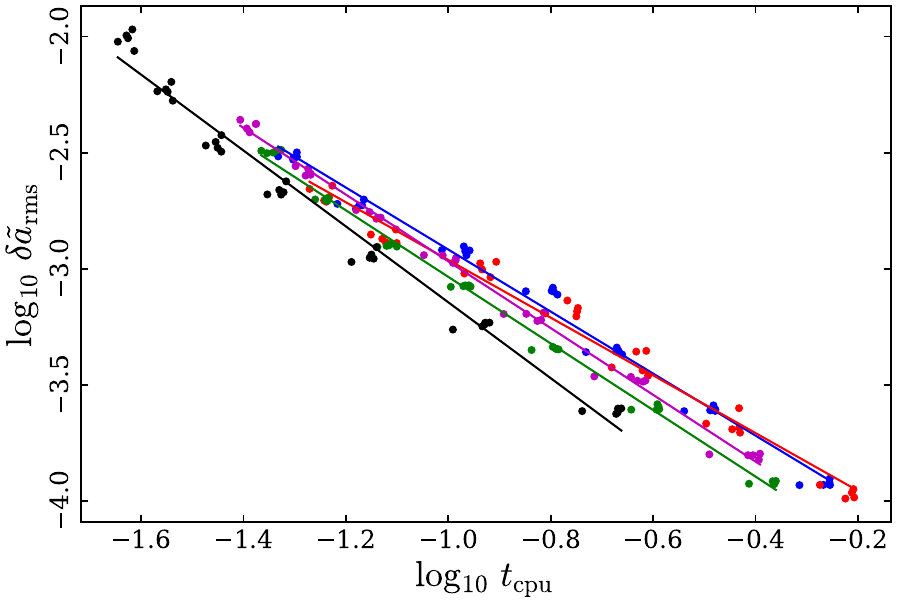}
  \caption{Acceleration error versus computing time.}
  \label{fig:scat_accrms_CPU}
\end{figure}